\documentclass[acmsmall]{acmart}

\usepackage[hyphenbreaks]{breakurl} 

\usepackage[english]{babel}
\usepackage{setspace}
\usepackage{calc}

\usepackage{paralist}
\usepackage{tikz}
\usetikzlibrary{arrows}

\usepackage{isabelle,isabellesym}
\isabellestyle{it}

\renewenvironment{isabelle}{%
  \medbreak\noindent\hspace{\parindent}%
  \begin{minipage}{\textwidth-\parindent}
  \begin{isabellebody}%
  \begin{tabbing}%
}{%
  \end{tabbing}%
  \end{isabellebody}%
  \end{minipage}%
  \medbreak%
}

\setcopyright{rightsretained}
\acmPrice{}
\acmDOI{10.1145/3133933}
\acmYear{2017}
\copyrightyear{2017}
\acmJournal{PACMPL}
\acmVolume{1}
\acmNumber{OOPSLA}
\acmArticle{109}
\acmMonth{10}

\begin{document}
\title{Verifying Strong Eventual Consistency in Distributed Systems}

\author{Victor B.\ F.\ Gomes}
\orcid{0000-0002-2954-4648}
\affiliation{
  \position{Research Associate}
  \department{Computer Laboratory}
  \institution{University of Cambridge}
  \streetaddress{15 JJ Thomson Avenue}
  \city{Cambridge}
  \postcode{CB3 0FD}
  \country{UK}
}
\email{vb358@cam.ac.uk}

\author{Martin Kleppmann}
\orcid{0000-0001-7252-6958}
\affiliation{
  \position{Research Associate}
  \department{Computer Laboratory}
  \institution{University of Cambridge}
  \streetaddress{15 JJ Thomson Avenue}
  \city{Cambridge}
  \postcode{CB3 0FD}
  \country{UK}
}
\email{mk428@cam.ac.uk}

\author{Dominic P.\ Mulligan}
\orcid{0000-0003-4643-3541}
\affiliation{
  \position{Research Associate}
  \department{Computer Laboratory}
  \institution{University of Cambridge}
  \streetaddress{15 JJ Thomson Avenue}
  \city{Cambridge}
  \postcode{CB3 0FD}
  \country{UK}
}
\email{dpm36@cam.ac.uk}

\author{Alastair R.\ Beresford}
\orcid{0000-0003-0818-6535}
\affiliation{
  \position{Senior Lecturer}
  \department{Computer Laboratory}
  \institution{University of Cambridge}
  \streetaddress{15 JJ Thomson Avenue}
  \city{Cambridge}
  \postcode{CB3 0FD}
  \country{UK}
}
\email{arb33@cam.ac.uk}

\thanks{Authors' address: Computer Laboratory, 15 JJ Thomson Avenue, Cambridge, CB3 0FD, United Kingdom.}

\begin{abstract}
Data replication is used in distributed systems to maintain up-to-date copies of shared data across multiple computers in a network.
However, despite decades of research, algorithms for achieving consistency in replicated systems are still poorly understood.
Indeed, many published algorithms have later been shown to be incorrect, even some that were accompanied by supposed mechanised proofs of correctness.
In this work, we focus on the correctness of Conflict-free Replicated Data Types (CRDTs), a class of algorithm that provides strong eventual consistency guarantees for replicated data.
We develop a modular and reusable framework in the Isabelle/HOL interactive proof assistant for verifying the correctness of CRDT algorithms.
We avoid correctness issues that have dogged previous mechanised proofs in this area by including a network model in our formalisation, and proving that our theorems hold in all possible network behaviours.
Our axiomatic network model is a standard abstraction that accurately reflects the behaviour of real-world computer networks.
Moreover, we identify an abstract convergence theorem, a property of order relations, which provides a formal definition of strong eventual consistency.
We then obtain the first machine-checked correctness theorems for three concrete CRDTs: the Replicated Growable Array, the Observed-Remove Set, and an Increment-Decrement Counter.
We find that our framework is highly reusable, developing proofs of correctness for the latter two CRDTs in a few hours and with relatively little CRDT-specific code.
\end{abstract}



\maketitle


\section{Introduction}
\label{sect.introduction}

A data replication algorithm is executed by a set of computers---or \emph{nodes}---in a distributed system, and ensures that all nodes eventually obtain an identical copy of some shared state.
Whilst vital for overall systems correctness, implementing a replication algorithm is a challenging task, as any such algorithm must operate across computer networks that may arbitrarily delay, drop, or reorder messages, experience temporary partitions of the nodes, or even suffer outright node failure.
Reflecting the importance of this task, a number of replication algorithms exist, with different algorithms exploring the inherent trade-offs between the strength of data consistency guarantees, and operational characteristics such as scalability and performance.
Accordingly, replication algorithms can be divided into classes---\emph{strong consistency}, \emph{eventual consistency}, and \emph{strong eventual consistency}---based on the consistency guarantees that they provide.

Strong consistency can be understood as \emph{linearisability}, \emph{serialisability}, or a combination of the two (one-copy serialisability).
Informally, the goal of strong consistency is to make a system behave like a single sequentially executing node, even when it is replicated and concurrent.
Most systems implement strong consistency by designating a single node as the \emph{leader}, which decides on a total order of operations and prevents concurrent access from causing conflicts.
Many relational databases, such as PostgreSQL, use this model.

However, strong consistency may be unwarranted or unnecessary depending on the application: it may impose an unacceptable performance degradation on the system, or it may simply be unfeasible to implement, especially in large distributed systems.
Relying on a single leader or central server limits the use and deployment of these systems: the server may become a bottleneck that limits scalability, and it makes the system vulnerable to disruption by network outages, denial-of-service attacks, censorship, and server failures.
Clients must constantly communicate with the leader in order to perform operations; if a node cannot reach the leader due to a network fault, its execution is stalled.
This fact makes strong consistency unsuitable for mobile devices, such as laptops and smartphones, that have intermittent network connectivity and must work offline.
It also rules out approaches that bypass the central server by using a local network for replication.

By contrast, decentralised or peer-to-peer architectures with weaker consistency models are able to provide better performance, fault-tolerance, and scalability characteristics.
One widely-implemented model is \emph{eventual consistency}, which guarantees that if no new updates are made to the shared state, all nodes will eventually have the same data \cite{Bailis:2013jc,Burckhardt:2014hy,Terry:1994fp,Vogels:2009ca}.
Since this model allows conflicting updates to be made concurrently, it requires a mechanism for resolving such conflicts.
For example, version control systems such as Git or Mercurial require the user to resolve merge conflicts manually; and some ``NoSQL'' distributed database systems such as Cassandra adopt a \emph{last-writer-wins} policy, under which one update is chosen as the winner, and concurrent updates are discarded \cite{KingsburyCassandra}.
Eventual consistency offers weak guarantees: it does not constrain the system behaviour when updates never cease, or the values that read operations may return prior to convergence.

\emph{Strong eventual consistency} (SEC) is a model that strikes a compromise between strong and eventual consistency~\cite{Shapiro:2011un}.
Informally, it guarantees that whenever two nodes have received the same set of updates---possibly in a different order---their view of the shared state is identical, and any conflicting updates are merged automatically.
Large-scale deployments of SEC algorithms include datacentre-based applications using Riak \cite{Brown:2014hs}, and collaborative editing applications such as Google Docs \cite{DayRichter:2010tt}.

Unlike strong consistency models, it is possible to implement SEC in decentralised settings without any central server or leader, and it allows local execution at each node to proceed without waiting for communication with other nodes.
However, algorithms for achieving decentralised SEC are currently poorly understood: several such algorithms, published in peer-reviewed venues, were subsequently shown to violate their supposed guarantees \cite{Imine:2003ks,Imine:2006kn,Oster:2005vi}.
As we show in Section~\ref{sect.relatedwork}, informal reasoning has repeatedly produced plausible-looking but incorrect algorithms, and there have even been examples of mechanised formal proofs of SEC algorithm correctness later being shown to be flawed \cite{Oster:2005vi}.
These mechanised proofs failed because, in formalising the algorithm, they made false assumptions about the execution environment.

In this work we use the Isabelle/HOL proof assistant~\cite{DBLP:conf/tphol/WenzelPN08} to create a framework for reliably reasoning about the correctness of a particular class of decentralised replication algorithms.
We do this by formalising not only the replication algorithms, but also the network in which they execute, allowing us to prove that the algorithm's assumptions hold in all possible network behaviours.
We model the network using the axioms of \emph{asynchronous unreliable causal broadcast}, a well-understood abstraction that is commonly implemented by network protocols, and which can run on almost any computer network, including large-scale networks that delay, reorder, or drop messages, and in which nodes may fail.

We then use this framework to produce machine-checked proofs of correctness for three Conflict-Free Replicated Data Types (CRDTs), a class of replication algorithms that ensure strong eventual consistency \cite{Shapiro:2011wy,Shapiro:2011un}.
These algorithms are suitable for use on mobile devices, which are not always connected to the Internet, but which may have a local connection (e.g. via Bluetooth) to other nodes carrying copies of the shared state.
We have used these algorithms to build a collaborative text editing application, and we plan to encapsulate them in a library that will allow developers to easily build applications that require data synchronisation, such as collaboratively editable spreadsheets, shared calendars, address books, and note-taking tools.

Our contributions in this paper are as follows:
\begin{itemize}
\item
We establish a framework for proving the strong eventual consistency (SEC) property of replication algorithms.
Our approach is ``foundational'' in the sense that we start with a general-purpose model of asynchronous unreliable causal broadcast networks---a communication abstraction that is compatible with virtually all network technologies today---and build up composable layers towards a full proof of correctness for a particular algorithm.
To our knowledge, this is the first machine-checked verification of SEC algorithms that explicitly models the network and reasons about all possible network behaviours.
The framework is modular and reusable, making it easy to formulate proofs for new algorithms.
\item
We provide the first mechanised proofs of correctness for the Replicated Growable Array (RGA), the operation-based Observed-Remove Set, and the operation-based counter CRDT.
RGA is an especially subtle algorithm: \citet{Attiya:2016kh} wrote, ``the reason why RGA actually works has been a bit of a mystery'', making its formal verification of interest, whilst the ORSet is supported as a primitive by the Lasp language~\cite{DBLP:conf/ppdp/MeiklejohnR15} for synchronisation-free programming, with an implemention also exported by the Akka framework~\cite{akka}.
These proofs demonstrate that our framework is highly reusable: we were able to quickly develop proofs of convergence for the set and counter CRDTs with little CRDT-specific code, using a fixed proof pattern that applies to all of our CRDTs.
All of our CRDT implementations are ``executable'' in the sense that functioning OCaml (or Scala, SML, and Haskell) code can be obtained from our definitions using Isabelle's code generation mechanism, and in experiments we have used one extracted implementation, sitting above a simple TCP network of $n$ nodes, to show that our implementations are usable in practice.
\item
As part of our proof framework, we identify an abstract convergence theorem, a property of order relations, from which we can deduce correctness theorems for concrete SEC algorithms.
Intuitively, this theorem can be viewed as the ``essence'' of why strong eventual consistency algorithms converge.
The convergence theorems for our three concrete CRDTs are obtained as direct corollaries of this theorem.
\end{itemize}

Our Isabelle theory files are open source\footnote{\url{https://github.com/trvedata/crdt-isabelle}} and included in the Archive of Formal Proofs \cite{CRDT-AFP}, enabling others to build upon our proof framework.

\section{High-level proof strategy}
\label{sect.high-level.proof.strategy}

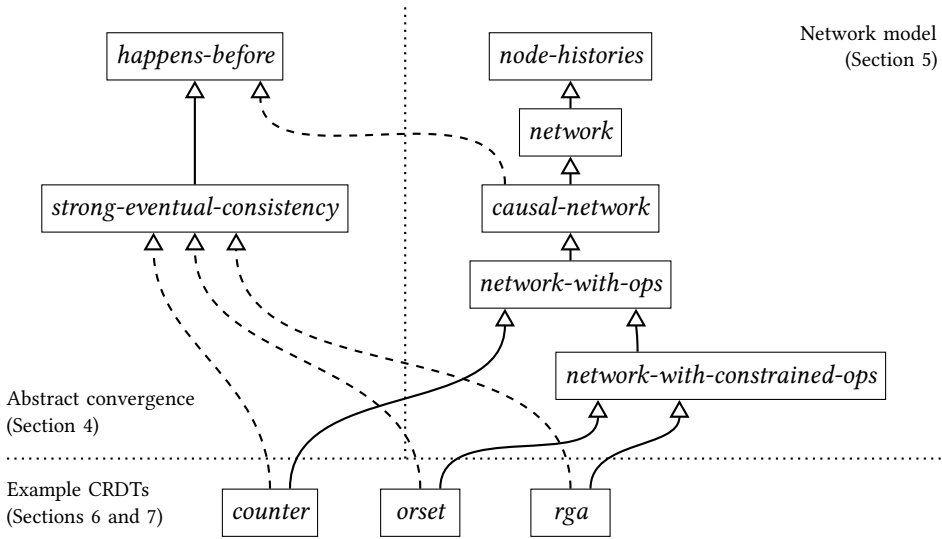
\begin{figure}
\centering
\begin{tikzpicture}[auto,scale=1.0]

\tikzstyle{locale}=[draw,anchor=base,minimum width=30pt,text height=8pt,text depth=3pt]

\node (hb)       at (0,6)   [locale] {$\isa{happens-before}$};
\node (sec)      at (0,4)   [locale] {$\isa{strong-eventual-consistency}$};
\node (hist)     at (5,6)   [locale] {$\isa{node-histories}$};
\node (network)  at (5,5)   [locale] {$\isa{network}$};
\node (causal)   at (5,4)   [locale] {$\isa{causal-network}$};
\node (netops)   at (5,3)   [locale] {$\isa{network-with-ops}$};
\node (netops2)  at (7,1.8) [locale] {$\isa{network-with-constrained-ops}$};
\node (counter)  at (1,0)   [locale] {$\isa{counter}$};
\node (orset)    at (3,0)   [locale] {$\isa{orset}$};
\node (rga)      at (5,0)   [locale] {$\isa{rga}$};

\begin{scope}[thick,dotted]
    \path [draw] (-2.5,0.8) -- ( 10,0.8);
    \path [draw] ( 2.8,0.8) -- (2.8,6.8);
\end{scope}

\begin{scope}[left,text width=5cm,text ragged left,font=\footnotesize]
    \node at (10, 6.2) {Network model\\(Section \ref{sect.network})};
\end{scope}
\begin{scope}[text width=5cm,font=\footnotesize]
    \node at (0, 1.4) {Abstract convergence\\(Section \ref{sect.abstract.convergence})};
    \node at (0, 0.2) {Example CRDTs\\(Sections \ref{sect.rga} and \ref{sect.simple.crdts})};
\end{scope}

\begin{scope}[>=open triangle 60]
    \tikzstyle{every path}=[thick,->]
    \draw (sec) to (hb);
    \draw (network) to (hist);
    \draw (causal) to (network);
    \draw (netops) to (causal);
    \draw (node cs:name=netops2,angle=164) to [out=90,in=270] (node cs:name=netops, angle=340);
    \draw (node cs:name=counter,angle=50)  to [out=90,in=270] (node cs:name=netops, angle=200);
    \draw (node cs:name=orset,  angle=50)  to [out=90,in=270] (node cs:name=netops2,angle=191);
    \draw (node cs:name=rga,    angle=50)  to [out=90,in=270] (node cs:name=netops2,angle=210);
    \tikzstyle{every path}=[thick,dashed,->]
    \draw (node cs:name=causal, angle=160) to [out=90,in=270] (node cs:name=hb, angle=340);
    \draw (node cs:name=counter,angle=90)  to [out=90,in=270] (node cs:name=sec,angle=210);
    \draw (node cs:name=orset,  angle=90)  to [out=90,in=270] (node cs:name=sec,angle=270);
    \draw (node cs:name=rga,    angle=90)  to [out=90,in=270] (node cs:name=sec,angle=330);
\end{scope}
\end{tikzpicture}
\caption{The main locales (modules) of our proof, and the relationships between them.
Solid arrows indicate a more specialised locale that extends a more general locale (like extending interfaces in OOP).
Dashed arrows indicate a sublocale that satisfies the assumptions of the superlocale (like implementing an interface in OOP).
}\label{fig.proof.structure}
\end{figure}

Since our formalisation of distributed algorithms goes into greater depth than prior work on strong eventual consistency, it is important to have a structure that keeps the proofs manageable.
Our approach breaks the proof into simple modules with cleanly defined properties---called \emph{locales}, a standard sectioning mechanism of Isabelle/HOL that will be described in Section~\ref{subsect.an.overview.of.isabelle} below---and composes them in order to describe more complex objects.
This locale structure is illustrated in Figure~\ref{fig.proof.structure} and explained below.

By lines of code, more than half of our proof is used to construct a general-purpose model of consistency in distributed systems, described in Section~\ref{sect.abstract.convergence}, and an axiomatic model of a computer network, described in Section~\ref{sect.network}, with both modules independent of any particular replication algorithm.
The remainder describes a formalisation of three CRDTs and their proofs of correctness, described in Sections~\ref{sect.rga} and~\ref{sect.simple.crdts}.
By keeping the general-purpose modules abstract and implementation-independent, we construct a reusable library of specifications and theorems.

We describe our formalisation of strong eventual consistency in Section~\ref{sect.abstract.convergence}.
In particular, we define what we mean by convergence, and prove an \emph{abstract convergence theorem}, which shows that the state of nodes converges if concurrent operations commute.
We are able to prove this fact without mentioning networks or any particular CRDT, but merely by reasoning about the ordering and properties of operations.
This definition constitutes a formal specification of what we mean by strong eventual consistency.

In Section~\ref{sect.network} we describe an axiomatic model of asynchronous networks.
The definition of the network is important because it allows us to prove that the desired properties hold in \emph{all} possible network behaviours, and that we are not making any dangerous assumptions that might be violated---an aspect that has dogged previous verification efforts for related algorithms (see Section~\ref{sect.related.verification}).
The network is the only part of our proof in which we make any axiomatic assumptions, and we show in Section~\ref{sect.network} that our assumptions are realistic, reflecting both standard conventions for modelling distributed systems, and the practical realities of network protocols today.
We then prove that our network satisfies the ordering properties required by the abstract convergence theorem of Section~\ref{sect.abstract.convergence}, and thus deduce a convergence theorem for our network model.

We use the general-purpose theorems and definitions from Sections~\ref{sect.abstract.convergence} and~\ref{sect.network} to prove the strong eventual consistency properties of concrete algorithms.
In Section~\ref{sect.rga} we describe our formalisation of the Replicated Growable Array (RGA), a CRDT for ordered lists.
We first show how to implement the RGA's insert and delete operations, with proofs that each operation commutes with itself, and that all operations commute with each other.
Insertion and deletion only commute under various conditions, so we prove that these conditions are satisfied in all possible network behaviours, and thus we obtain a concrete convergence theorem for our RGA implementation.
Next, in Section~\ref{sect.simple.crdts}, we demonstrate the generality of our proof framework with definitions of two simple CRDTs: a Counter and an Observed-Remove Set.

As illustrated in Figure~\ref{fig.proof.structure}, the $\isa{counter}$, $\isa{orset}$, and $\isa{rga}$ locales can use the definitions and lemmas of the network model because they extend that model.
We then prove that all three locales satisfy the abstract specification $\isa{strong-eventual-consistency}$, and therefore show that these algorithms provide strong eventual consistency.

\section{An Introduction to Isabelle}
\label{subsect.an.overview.of.isabelle}

We now provide a brief introduction to the key concepts and syntax of Isabelle/HOL.
Familiar readers may skip to Section~\ref{sect.abstract.convergence}.
A more detailed introduction can be found in the standard tutorial material~\cite{DBLP:books/sp/NipkowK14}.

\paragraph{Syntax of expressions.}

Isabelle/HOL is a logic with a strict, polymorphic, inferred type system.
\emph{Function types} are written $\tau_1 \Rightarrow \tau_2$, and are inhabited by \emph{total} functions, mapping elements of $\tau_1$ to elements of $\tau_2$.
We write $\tau_1 \times \tau_2$ for the \emph{product type} of $\tau_1$ and $\tau_2$, inhabited by pairs of elements of type $\tau_1$ and $\tau_2$, respectively.
In a similar fashion to Standard ML and OCaml, \emph{type operators} are applied to arguments in reverse order, and therefore $\tau\ \isa{list}$ denotes the type of lists of elements of type $\tau$, and $\tau\ \isa{set}$ denotes the type of mathematical (i.e., potentially infinite) sets of type $\tau$.
Type variables are written in lowercase, and preceded with a prime: ${\isacharprime}a \Rightarrow {\isacharprime}a$ denotes the type of a polymorphic identity function, for example.
\emph{Tagged union} types are introduced with the $\isacommand{datatype}$ keyword, with constructors of these types usually written with an initial upper case letter.

In Isabelle/HOL's term language we write $\isa{t} \mathbin{::} \tau$ for a \emph{type ascription}, constraining the type of the term $\isa{t}$ to the type $\tau$.
We write $\lambda{x}.\: t$ for an anonymous function mapping an argument $\isa{x}$ to $\isa{t(x)}$, and write the application of term $\isa{t}$ with function type to an argument $\isa{u}$ as $\isa{t\ u}$, as usual.
Terms of list type are introduced using one of two constructors: the empty list $[\,]$ or `nil', and the infix operator $\isa{\#}$ which is pronounced ``cons'', and which prepends an element to an existing list.
We use $[t_1, \ldots, t_n]$ as syntactic sugar for a list literal, and $\isa{xs} \mathbin{\isacharat} \isa{ys}$ to express the concatenation (appending) of two lists $\isa{xs}$ and $\isa{ys}$.
We write $\{\,\}$ for the empty set, and use usual mathematical notation for set union, disjunction, membership tests, and so on: $\isa{t} \cup \isa{u}$, $\isa{t} \cap \isa{u}$, and $\isa{x} \in \isa{t}$.
We write $t \longrightarrow s$ for logical implication between formulae (terms of type $\isa{bool}$).
Strictly speaking Isabelle is a logical framework, providing a weak meta-logic within which object logics are embedded, including the Isabelle/HOL object logic that we use in this work.
Accordingly, the implication arrow of Isabelle's meta-logic, $\isa{t} \Longrightarrow \isa{u}$, is required in certain contexts over the object-logic implication arrow, $t \longrightarrow s$, already introduced.
However, for purposes of an intuitive understanding, the two forms of implication can be regarded as equivalent by the reader, with the requirement to use one over the other merely being an implementation detail of Isabelle itself.
We will sometimes use the shorthand ${\isasymlbrakk}\isa{H}_1{\isacharsemicolon}\ \ldots{\isacharsemicolon}\ \isa{H}_n{\isasymrbrakk}\ {\isasymLongrightarrow}\ C$ instead of iterated meta-logic implications, i.e., $H_1\ {\isasymLongrightarrow}\ \ldots\ {\isasymLongrightarrow}\ H_n\ {\isasymLongrightarrow}\ C$.

\paragraph{Definitions and theorems.}

New non-recursive definitions are entered into Isabelle's global context using the $\mathbf{definition}$ keyword.
Recursive functions are defined using the $\mathbf{fun}$ keyword, and support pattern matching on their arguments.
All functions are total, and therefore every recursive function must be provably terminating.
The termination proofs in this work are generated automatically by Isabelle itself.

Inductive relations are defined with the $\mathbf{inductive}$ keyword.
For example, the definition
\begin{isabelle}
\isacommand{inductive} only-fives\ {\isacharcolon}{\isacharcolon}\ {\isachardoublequoteopen}nat\ list\ {\isasymRightarrow}\ bool{\isachardoublequoteclose}\ \isakeyword{where}\\
~~~~{\isachardoublequoteopen}only-fives\ {\isacharbrackleft}{\isacharbrackright}{\isachardoublequoteclose}\ {\isacharbar}\\
~~~~{\isachardoublequoteopen}{\isasymlbrakk}\ only-fives\ xs\ {\isasymrbrakk}\ {\isasymLongrightarrow}\ only-fives {\isacharparenleft}5\#xs{\isacharparenright}{\isachardoublequoteclose}
\end{isabelle}
\noindent
introduces a new constant $\isa{only-fives}$ of type $\isa{nat list} \Rightarrow \isa{bool}$.
The two clauses in the body of the definition enumerate the conditions under which $\isa{only-fives}\ \isa{xs}$ is true, for arbitrary $\isa{xs}$: firstly, $\isa{only-fives}$ is true for the empty list; and secondly, if you know that $\isa{only-fives}\ \isa{xs}$ is true for some $\isa{xs}$, then you can deduce that $\isa{only-fives}\ (5\#\isa{xs})$ (i.e., $\isa{xs}$ prefixed with the number 5) is also true.
Moreover, $\isa{only-fives}\ \isa{xs}$ is true in no other circumstances---it is the \emph{smallest} relation closed under the rules defining it.
In short, the clauses above state that $\isa{only-fives}\ \isa{xs}$ holds exactly in the case where $\isa{xs}$ is a (potentially empty) list containing only repeated copies of the natural number $5$.

Lemmas, theorems, and corollaries can be asserted using the $\isacommand{lemma}$, $\isacommand{theorem}$, and $\isacommand{corollary}$ keywords, respectively.
There is no semantic difference between these keywords in Isabelle.
For example,
\begin{isabelle}
~~~~\isakeyword{assumes}\ \=\kill
\isacommand{theorem} only-fives-concat{\isacharcolon}\\
~~~~\isakeyword{assumes}\>only-fives\ xs \isakeyword{and}\ only-fives\ ys\\
~~~~\isakeyword{shows}\>only-fives (xs \isacharat ys)
\end{isabelle}
\noindent
conjectures that if $\isa{xs}$ and $\isa{ys}$ are both lists of fives, then their concatenation $xs \mathbin{\isacharat} ys$ is also a list of fives.
Isabelle then requires that this claim be proved by using one of its proof methods, for example by induction.
Some proofs can be automated, whilst others require the user to provide explicit reasoning steps.
The theorem is assigned a name, here $\isa{only-fives-concat}$, so that it may be referenced in later proofs.

\paragraph{Locales.}

Lastly, we use \emph{locales}---or local theories~\cite{DBLP:conf/tphol/KammullerWP99,DBLP:conf/types/HaftmannW08}---extensively to structure the proof, as shown in Figure~\ref{fig.proof.structure}.
In programming terms, Isabelle's locales may be thought of as an interface with associated laws that implementations must obey.
In particular, a declaration of the form
\begin{isabelle}
~~~~\isakeyword{assumes}\ \=\kill
\isacommand{locale} semigroup =\\
~~~~\isakeyword{fixes}\>f\ {\isacharcolon}{\isacharcolon}\ {\isachardoublequoteopen}{\isacharprime}a\ {\isasymRightarrow}\ {\isacharprime}a{\isachardoublequoteclose}\ {\isasymRightarrow}\ {\isacharprime}a{\isachardoublequoteclose}\\
~~~~\isakeyword{assumes}\>{\isachardoublequoteopen}f\ x\ (f\ y\ z)\ =\ f\ (f\ x\ y)\ z{\isachardoublequoteclose}
\end{isabelle}
\noindent
introduces a locale, with a fixed, typed constant $\isa{f}$, and a law asserting that $\isa{f}$ is associative.
Functions and constants may now be defined, and theorems conjectured and proved, within the context of the $\isa{semigoup}$ locale, i.e. definitions may be made ``generic'' in a semigroup.
This is indicated syntactically by writing $(\isacommand{in}\ \isa{semigroup})$ before the name of the constant being defined, or the theorem being conjectured, at the point of definition or conjecture.
Any function, constant, or theorem, marked in this way may make reference to $\isa{f}$, or the fact that $\isa{f}$ is associative.
\emph{Interpreting} a locale---such as $\isa{semigroup}$ above---involves providing a concrete implementation of $\isa{f}$ coupled with a proof that the concrete implementation satisfies the associated law, and is akin to implementing an interface.
Once interpreted, all functions, definitions, and theorems made within the $\isa{semigroup}$ locale become available to use for that concrete implementation.
Like interfaces, locales may be extended with new functionality, and may be specialised, by other ``sublocales'', forming a hierarchy.

\section{Abstract convergence}
\label{sect.abstract.convergence}

Strong eventual consistency (SEC) requires \emph{convergence} of all copies of the shared state: whenever two nodes have received the same set of updates, they must be in the same state.
This definition constrains the values that read operations may return at any time, making SEC a stronger property than eventual consistency.
By accessing only their local copy of the shared state, nodes can execute read and write operations without waiting for network communication.
Nodes exchange updates asynchronously when a network connection is available.  

We now use Isabelle to formalise the notion of strong eventual consistency.
In this section we do not make any assumptions about networks or data structures; instead, we use an abstract model of operations that may be reordered, and we reason about the properties that those operations must satisfy.
We then provide concrete implementations of that abstract model in later sections.

\subsection{The happens-before relation and causality}\label{sect.happens.before}

The simplest way of achieving convergence is to require all operations to be commutative, but this definition is too strong to be useful for many datatypes.
For example, in a set, an element may first be added and then subsequently removed again.
Although it is possible to make such additions and removals unconditionally commutative, doing so yields counter-intuitive semantics \cite{Bieniusa:2012wu,Bieniusa:2012gt}.
Instead, a better approach is to require only \emph{concurrent} operations to commute with each other.
Two operations are concurrent if neither ``knew about'' the other at the time when they were generated.
If one operation happened before another---for example, if the removal of an element from a set knew about the prior addition of that element from the set---then it is reasonable to assume that all nodes will apply the operations in that order (first the addition, then the removal).

The \emph{happens-before} relation, as introduced by \citet{Lamport:1978jq}, captures such causal dependencies between operations.
It can be defined in terms of sending and receiving messages on a network, and we give such a definition in Section~\ref{sect.network}.
However, for now, we keep it abstract, writing $\isa{x} \prec \isa{y}$ to indicate that operation $\isa{x}$ happened before $\isa{y}$, where $\prec$ is a predicate of type $\isacharprime\isa{oper} \mathbin{\isasymRightarrow} \isacharprime\isa{oper} \mathbin{\isasymRightarrow} \isa{bool}$.
In words, $\prec$ can be applied to two operations of some abstract type $\isacharprime\isa{oper}$, returning either $\isa{True}$ or $\isa{False}$.%
\footnote{Note that in the distributed systems literature it is conventional to write the happens-before relation as $\isa{x} \rightarrow \isa{y}$, but we reserve the arrow operator to denote logical implication.}
Our only restriction on the happens-before relation $\prec$ is that it must be a \emph{strict partial order}, that is, it must be irreflexive and transitive, which implies that it is also antisymmetric.
We say that two operations $\isa{x}$ and $\isa{y}$ are \emph{concurrent}, written $\isa{x} \mathbin{\isasymparallel} \isa{y}$, whenever one does not happen before the other:
$\neg (\isa{x} \prec \isa{y})$ and $\neg (\isa{y} \prec \isa{x})$.
Thus, given any two operations $\isa{x}$ and $\isa{y}$, there are three mutually exclusive ways in which they can be related: either $\isa{x} \prec \isa{y}$, or $\isa{y} \prec \isa{x}$, or $\isa{x} \mathbin{\isasymparallel} \isa{y}$.

As discussed above, the purpose of the happens-before relation is to require that some operations must be applied in a particular order, while allowing concurrent operations to be reordered with respect to each other.
We assume that each node applies operations in some sequential order (a standard assumption for distributed algorithms), and so we can model the execution history of a node as a list of operations.
We can then inductively define a list of operations as being \emph{consistent with the happens-before relation}, or simply \emph{hb-consistent}, as follows:
\begin{isabelle}
\isacommand{inductive} hb{\isacharunderscore}consistent\ {\isacharcolon}{\isacharcolon}\ {\isachardoublequoteopen}{\isacharprime}oper\ list\ {\isasymRightarrow}\ bool{\isachardoublequoteclose}\ \isakeyword{where}\\
~~~~{\isachardoublequoteopen}hb{\isacharunderscore}consistent\ {\isacharbrackleft}{\isacharbrackright}{\isachardoublequoteclose}\ {\isacharbar}\\
~~~~{\isachardoublequoteopen}{\isasymlbrakk}\ hb{\isacharunderscore}consistent\ xs{\isacharsemicolon}\ {\isasymforall}x\ {\isasymin}\ set\ xs{\isachardot}\ {\isasymnot}\ y\ {\isasymprec}\ x\ {\isasymrbrakk}\ {\isasymLongrightarrow}\ hb{\isacharunderscore}consistent\ {\isacharparenleft}xs\ {\isacharat}\ {\isacharbrackleft}y{\isacharbrackright}{\isacharparenright}{\isachardoublequoteclose}
\end{isabelle}
In words: the empty list is hb-consistent; furthermore, given an hb-consistent list $\isa{xs}$, we can append an operation $\isa{y}$ to the end of the list to obtain another hb-consistent list, provided that $\isa{y}$ does not happen-before any existing operation $\isa{x}$ in $\isa{xs}$. As a result, whenever two operations $\isa{x}$ and $\isa{y}$ appear in a hb-consistent list, and $\isa{x}\prec\isa{y}$, then $\isa{x}$ must appear before $\isa{y}$ in the list. However, if $\isa{x}\mathbin{\isasymparallel}\isa{y}$, the operations can appear in the list in either order.

\subsection{Interpretation of operations}\label{sect.ops.interpretation}

We describe the state of a node using an abstract type variable $\isacharprime\isa{state}$.
To model state changes, we assume the existence of an \emph{interpretation} function of type $\isa{interp} \mathbin{\isacharcolon\isacharcolon} \isacharprime\isa{oper} \mathbin{\isasymRightarrow} \isacharprime\isa{state} \mathbin{\isasymRightarrow} \isacharprime\isa{state}\ \isa{option}$, which lifts an operation into a \emph{state transformer}---a function that either maps an old state to a new state, or fails by returning $\isa{None}$.
If $\isa{x}$ is an operation, we also write $\langle\isa{x}\rangle$ for the state transformer obtained by applying $\isa{x}$ to the interpretation function.

Concretely, these definitions are captured in Isabelle with the following locale declaration:
\begin{isabelle}
~~~~\isakeyword{fixes}\ \=hb{\isacharunderscore}weak\ \=\kill
\isacommand{locale} happens{\isacharunderscore}before\ {\isacharequal}\ preorder\ hb{\isacharunderscore}weak\ hb\\
~~~~\isakeyword{for}\>hb{\isacharunderscore}weak\>{\isacharcolon}{\isacharcolon}\ {\isachardoublequoteopen}{\isacharprime}oper\ {\isasymRightarrow}\ {\isacharprime}oper\ {\isasymRightarrow}\ bool{\isachardoublequoteclose}\\
~~~~\isakeyword{and}\>hb\>{\isacharcolon}{\isacharcolon}\ {\isachardoublequoteopen}{\isacharprime}oper\ {\isasymRightarrow}\ {\isacharprime}oper\ {\isasymRightarrow}\ bool{\isachardoublequoteclose}\ {\isacharplus}\\
~~~~\isakeyword{fixes}\>interp\>{\isacharcolon}{\isacharcolon}\ {\isachardoublequoteopen}{\isacharprime}oper\ {\isasymRightarrow}\ {\isacharprime}state\ {\isasymRightarrow}\ {\isacharprime}state\ option{\isachardoublequoteclose}
\end{isabelle}
The $\isa{happens-before}$ locale extends the $\isa{preorder}$ locale, which is part of Isabelle's standard library and includes various useful lemmas.
It fixes two constants: a preorder that we call $\isa{hb-weak}$ or $\preceq$, and a strict partial order that we call $\isa{hb}$ or $\prec$.
We are only interested in the strict partial order and define $\isa{x}\preceq\isa{y}$ to be $\isa{x}\prec\isa{y} \vee \isa{x}=\isa{y}$.
Moreover, the locale fixes the interpretation function $\isa{interp}$ as described above, which means that we assume the existence of a function with the given type signature without specifying an implementation.

Given two operations $\isa{x}$ and $\isa{y}$, we can now define the composition of state transformers: we write $\langle\isa{x}\rangle \mathbin{\isasymrhd} \langle\isa{y}\rangle$ to denote the state transformer that first applies the effect of $\isa{x}$ to some state, and then applies the effect of $\isa{y}$ to the result.
If either $\langle\isa{x}\rangle$ or $\langle\isa{y}\rangle$ fails, the combined state transformer also fails.
The operator $\isasymrhd$ is a specialised form of the \emph{Kleisli arrow composition}, which we define as:
\begin{isabelle}
\isacommand{definition} kleisli\ {\isacharcolon}{\isacharcolon}\ {\isachardoublequoteopen}{\isacharparenleft}{\isacharprime}a\ {\isasymRightarrow}\ {\isacharprime}a\ option{\isacharparenright}\ {\isasymRightarrow}\ {\isacharparenleft}{\isacharprime}a\ {\isasymRightarrow}\ {\isacharprime}a\ option{\isacharparenright}\ {\isasymRightarrow}\ {\isacharparenleft}{\isacharprime}a\ {\isasymRightarrow}\ {\isacharprime}a\ option{\isacharparenright}{\isachardoublequoteclose}\ \isakeyword{where}\\
~~~~{\isachardoublequoteopen}f\ {\isasymrhd}\ g\ {\isasymequiv}\ {\isasymlambda}x{\isachardot}\ f\ x\ {\isasymbind}\ {\isacharparenleft}{\isasymlambda}y{\isachardot}\ g\ y{\isacharparenright}{\isachardoublequoteclose}
\end{isabelle}
\noindent Here, $\isasymbind$ is the \emph{monadic bind} operation, defined on the option type that we are using to implement partial functions.
We can now define a function $\isa{apply-operations}$ that composes an arbitrary list of operations into a state transformer.
We first map $\isa{interp}$ across the list to obtain a state transformer for each operation, and then collectively compose them using the Kleisli arrow composition combinator:
\begin{isabelle}
\isacommand{definition} apply{\isacharunderscore}operations\ {\isacharcolon}{\isacharcolon}\ {\isachardoublequoteopen}{\isacharprime}oper\ list\ {\isasymRightarrow}\ {\isacharprime}state\ {\isasymRightarrow}\ {\isacharprime}state\ option{\isachardoublequoteclose}\ \isakeyword{where}\\
~~~~{\isachardoublequoteopen}apply{\isacharunderscore}operations\ ops\ {\isasymequiv}\ foldl\ {\isacharparenleft}op\ {\isasymrhd}{\isacharparenright}\ Some\ {\isacharparenleft}map\ interp\ ops{\isacharparenright}{\isachardoublequoteclose}
\end{isabelle}
\noindent The result is a state transformer that applies the interpretation of each of the operations in the list, in left-to-right order, to some initial state.
If any of the operations fails, the entire composition returns $\isa{None}$.

\subsection{Commutativity and convergence}\label{sect.ops.commute}

We say that two operations $\isa{x}$ and $\isa{y}$ \emph{commute} whenever $\langle\isa{x}\rangle \mathbin{\isasymrhd} \langle\isa{y}\rangle = \langle\isa{y}\rangle \mathbin{\isasymrhd} \langle\isa{x}\rangle$, i.e. when we can swap the order of the composition of their interpretations without changing the resulting state transformer.
For our purposes, requiring that this property holds for \emph{all} pairs of operations is too strong.
Rather, the commutation property is only required to hold for operations that are concurrent, as captured in the next definition:
\begin{isabelle}
\isacommand{definition} concurrent{\isacharunderscore}ops{\isacharunderscore}commute\ {\isacharcolon}{\isacharcolon}\ {\isachardoublequoteopen}{\isacharprime}oper\ list\ {\isasymRightarrow}\ bool{\isachardoublequoteclose}\ \isakeyword{where}\\
~~~~{\isachardoublequoteopen}concurrent{\isacharunderscore}ops{\isacharunderscore}commute\ xs\ {\isasymequiv} {\isasymforall}x\ y{\isachardot}\ {\isacharbraceleft}x{\isacharcomma}\ y{\isacharbraceright}\ {\isasymsubseteq}\ set\ xs\ {\isasymlongrightarrow}\ x\ {\isasymparallel}\ y\ {\isasymlongrightarrow}\ {\isasymlangle}x{\isasymrangle}{\isasymrhd}{\isasymlangle}y{\isasymrangle}\ {\isacharequal}\ {\isasymlangle}y{\isasymrangle}{\isasymrhd}{\isasymlangle}x{\isasymrangle}{\isachardoublequoteclose}
\end{isabelle}
Given this definition, we can now state and prove our main theorem, $\isa{convergence}$.
This theorem states that two hb-consistent lists of distinct operations, which are permutations of each other and in which concurrent operations commute, have the same interpretation:
\begin{isabelle}
\isacommand{theorem} convergence{\isacharcolon}\\
~~~~\isakeyword{assumes}\ {\isachardoublequoteopen}set\ xs\ {\isacharequal}\ set\ ys{\isachardoublequoteclose}\ \isakeyword{and}\ {\isachardoublequoteopen}concurrent{\isacharunderscore}ops{\isacharunderscore}commute\ xs{\isachardoublequoteclose}\ \isakeyword{and}\ {\isachardoublequoteopen}concurrent{\isacharunderscore}ops{\isacharunderscore}commute\ ys{\isachardoublequoteclose}\\
~~~~~~~~\isakeyword{and}\ {\isachardoublequoteopen}distinct\ xs{\isachardoublequoteclose}\ \isakeyword{and}\ {\isachardoublequoteopen}distinct\ ys{\isachardoublequoteclose}\ \isakeyword{and}\ {\isachardoublequoteopen}hb{\isacharunderscore}consistent\ xs{\isachardoublequoteclose}\ \isakeyword{and}\ {\isachardoublequoteopen}hb{\isacharunderscore}consistent\ ys{\isachardoublequoteclose}\\
~~~~\isakeyword{shows}\ {\isachardoublequoteopen}apply{\isacharunderscore}operations\ xs\ {\isacharequal}\ apply{\isacharunderscore}operations\ ys{\isachardoublequoteclose}
\end{isabelle}
\noindent
A fully mechanised proof of this theorem can be found in our submission to the Archive of Formal Proofs \cite{CRDT-AFP}.
Although this theorem may seem ``obvious'' at first glance---commutativity allows the operation order to be permuted---it is more subtle than it seems.
The difficulty arises because operations may succeed when applied to some state, but fail when applied to another state (for example, attempting to delete an element that does not exist in the state).
We find it interesting that it is nevertheless sufficient for the definition of $\isa{concurrent-ops-commute}$ to be expressed only in terms of the Kleisli arrow composition, and without explicitly referring to the state.

\subsection{Formalising Strong Eventual Consistency}\label{sect.abstract.sec.spec}

Besides convergence, another required property of SEC is \emph{progress}: if one node issues a valid operation, and another node applies that operation, then it must not become stuck in an error state.
Although the type signature of the interpretation function allows operations to fail, we need to prove that such a failure never occurs in any $\isa{hb-consistent}$ network behaviour.
We capture this requirement in the $\isa{strong-eventual-consistency}$ locale:
\begin{isabelle}
~~~~\isakeyword{assumes}\ \=commutativity{\isacharcolon}\ \={\isasymlbrakk}\ \=op{\isacharunderscore}history{\isacharparenleft}xs{\isacharat}{\isacharbrackleft}x{\isacharbrackright}{\isacharparenright}\ \={\isasymrbrakk}\ \=\kill
\isacommand{locale}\ strong{\isacharunderscore}eventual{\isacharunderscore}consistency\ {\isacharequal}\ happens{\isacharunderscore}before\ {\isacharplus}\\
~~~~\isakeyword{fixes}\ op{\isacharunderscore}history\ {\isacharcolon}{\isacharcolon}\ {\isachardoublequoteopen}{\isacharprime}oper\ list\ {\isasymRightarrow}\ bool{\isachardoublequoteclose}\ \ \isakeyword{and}\ initial{\isacharunderscore}state\ {\isacharcolon}{\isacharcolon}\ {\isachardoublequoteopen}{\isacharprime}state{\isachardoublequoteclose}\\
~~~~\isakeyword{assumes}\>causality{\isacharcolon}\ \>{\isasymlbrakk}\>{\isachardoublequoteopen}op{\isacharunderscore}history\ xs\ \>{\isasymrbrakk}\ {\isasymLongrightarrow}\ hb{\isacharunderscore}consistent\ xs{\isachardoublequoteclose}\\
~~~~~~~~\isakeyword{and}\>distinctness{\isacharcolon}\ \>{\isasymlbrakk}\>{\isachardoublequoteopen}op{\isacharunderscore}history\ xs\ \>{\isasymrbrakk}\ {\isasymLongrightarrow}\ distinct\ xs{\isachardoublequoteclose}\\
~~~~~~~~\isakeyword{and}\>trunc{\isacharunderscore}history{\isacharcolon}\ \>{\isasymlbrakk}\>{\isachardoublequoteopen}op{\isacharunderscore}history{\isacharparenleft}xs{\isacharat}{\isacharbrackleft}x{\isacharbrackright}{\isacharparenright}\ \>{\isasymrbrakk}\ {\isasymLongrightarrow}\ op{\isacharunderscore}history\ xs{\isachardoublequoteclose}\\
~~~~~~~~\isakeyword{and}\>commutativity{\isacharcolon}\ \>{\isasymlbrakk}\>{\isachardoublequoteopen}op{\isacharunderscore}history\ xs\ \>{\isasymrbrakk}\ {\isasymLongrightarrow}\ concurrent{\isacharunderscore}ops{\isacharunderscore}commute\ xs{\isachardoublequoteclose}\\
~~~~~~~~\isakeyword{and}\>no{\isacharunderscore}failure{\isacharcolon}\ \>{\isasymlbrakk}\>{\isachardoublequoteopen}op{\isacharunderscore}history{\isacharparenleft}xs{\isacharat}{\isacharbrackleft}x{\isacharbrackright}{\isacharparenright};\\
\>\>\>apply{\isacharunderscore}operations\ xs\ initial{\isacharunderscore}state\ {\isacharequal}\ Some\ state\\
\>\>\>\>{\isasymrbrakk}\>{\isasymLongrightarrow}\ {\isasymlangle}x{\isasymrangle}\ state\ {\isasymnoteq}\ None{\isachardoublequoteclose}
\end{isabelle}
\noindent Here, $\isa{op-history}$ is an abstract predicate describing any valid operation history of some replication algorithm, encapsulating the assumptions of the $\isa{convergence}$ theorem ($\isa{concurrent-ops-commute}$, $\isa{distinct}$, and $\isa{hb-consistent}$).
This locale serves as a concise summary of the properties that we require in order to achieve SEC, and from these assumptions and the theorem above we easily obtain the two safety properties of SEC as theorems:
\begin{isabelle}
~~~~\isakeyword{assumes}\ \=\kill
\isacommand{theorem}\ sec{\isacharunderscore}convergence{\isacharcolon}\\
~~~~\isakeyword{assumes}\ \>{\isachardoublequoteopen}set\ xs\ {\isacharequal}\ set\ ys{\isachardoublequoteclose}\ \isakeyword{and}\ {\isachardoublequoteopen}op{\isacharunderscore}history\ xs{\isachardoublequoteclose}\ \isakeyword{and}\ {\isachardoublequoteopen}op{\isacharunderscore}history\ ys{\isachardoublequoteclose}\\
~~~~\isakeyword{shows}\ \>{\isachardoublequoteopen}apply{\isacharunderscore}operations\ xs\ {\isacharequal}\ apply{\isacharunderscore}operations\ ys{\isachardoublequoteclose}\\[4pt]
\isacommand{theorem}\ sec{\isacharunderscore}progress{\isacharcolon}\\
~~~~\isakeyword{assumes} \>{\isachardoublequoteopen}op{\isacharunderscore}history\ xs{\isachardoublequoteclose}\\
~~~~\isakeyword{shows} \>{\isachardoublequoteopen}apply{\isacharunderscore}operations\ xs\ initial{\isacharunderscore}state\ {\isasymnoteq}\ None{\isachardoublequoteclose}
\end{isabelle}

Thus, in order to prove SEC for some replication algorithm, we only need to show that the five assumptions of the $\isa{strong-eventual-consistency}$ locale are satisfied.
As we shall see in Section~\ref{sect.network}, the first three assumptions are satisfied by our network model, and do not require any algorithm-specific proofs.
For individual algorithms we only need to prove the $\isa{commutativity}$ and $\isa{no-failure}$ properties, and we show how to do this in Sections~\ref{sect.rga} and~\ref{sect.simple.crdts}.

Note that the $\isa{trunc-history}$ assumption requires that every prefix of a valid operation history is also valid.
This means that the convergence theorem holds at every step of the execution, not only at some unspecified time in the future (``eventually''), making SEC stronger than eventual consistency.

\section{An axiomatic network model}
\label{sect.network}

In this section we develop a formal definition of an \emph{asynchronous unreliable causal broadcast network}.
We choose this model because it satisfies the causal delivery requirements of many operation-based CRDTs \cite{Almeida:2015fc,Baquero:2014ed}.
Moreover, it is suitable for use in decentralised settings, as motivated in the introduction, since it does not require waiting for communication with a central server or a quorum of nodes.
Stronger consistency models do not have this property \cite{Attiya:2015dm,Davidson:1985hv}.

The \emph{causal} and \emph{broadcast} aspects of the model are explained in Sections~\ref{sect.network.broadcast} and~\ref{sect.network.causal}.
The \emph{asynchronous} aspect means that we make no timing assumptions: messages sent over the network may suffer unbounded delays before they are delivered, nodes may pause their execution for unbounded periods of time, and we require no clock synchronisation.
\emph{Unreliable} means that messages may never arrive at all, and nodes may fail permanently without warning.
Networks are known to exhibit these behaviours in practice \cite{Bailis:2014jx}, and replication algorithms must tolerate such failures.

This model provides a realistic setting in which we can embed various replication algorithms, and prove that they guarantee SEC in all possible behaviours of the network.
But it is also abstract enough to be able to model a wide range of scenarios: for example, if a user makes updates while offline, and the device re-synchronises when it is next online, we can simply model that interaction as very large network delay.
Our network model is defined using only six axioms, all of which are standard assumptions when modelling distributed systems, and which are satisfied by many systems in practice.
All theorems in this paper are derived from those axioms; in particular, we show that the causal delivery abstraction satisfies the strict partial ordering assumption of $\isa{hb-consistent}$ (Section~\ref{sect.happens.before}), allowing us to use the convergence theorem in any locales that extend the network.

\subsection{Modelling a distributed system}

We model a distributed system as an unbounded number of communicating nodes.
We assume nothing about the communication pattern of nodes---we assume only that each node is uniquely identified by a natural number, and that the flow of execution at each node consists of a finite, totally ordered sequence of execution steps (events).
We call that sequence of events at node $i$ the \emph{history} of that node.
For convenience, we assume that every event or execution step is unique within a node's history; this assumption is standard when modelling distributed systems \cite{Cachin:2011wt} and can easily be implemented by attaching a sequence number, timestamp, or other unique identifier to each event.
This system model can be expressed in Isabelle as follows:
\begin{isabelle}
~~~~\isakeyword{assumes}\ \=\kill
\isacommand{locale} node{\isacharunderscore}histories\ {\isacharequal}\\
~~~~\isakeyword{fixes}\ \>history\ {\isacharcolon}{\isacharcolon}\ {\isachardoublequoteopen}nat\ {\isasymRightarrow}\ {\isacharprime}a\ list{\isachardoublequoteclose}\ \\
~~~~\isakeyword{assumes}\ \>histories{\isacharunderscore}distinct{\isacharcolon}\ {\isachardoublequoteopen}distinct\ {\isacharparenleft}history\ i{\isacharparenright}{\isachardoublequoteclose}
\end{isabelle}
Here, the history of a node $\isa{i}$ is obtained by using a function fixed by the locale, $\isa{history}$.
The history is simply a list of events, and each event is modelled as an abstract type variable---here we use $\isa{{\isacharprime}a}$.
The $\isa{distinct}$ predicate is an Isabelle/HOL library function that asserts that a list contains no duplicate elements.
Note that we make no assumption about the number of nodes in the system, which allows us to model systems in which nodes join and leave the network over time.
A node that does not exist is simply modelled as an empty list of events.

A node's history is finite, and at the end of a node's history we assume that a node has either failed or successfully terminated.
We treat node failure as permanent, and model it by the absence of any further events in its history.
This \emph{crash-stop} abstraction is commonly used by distributed algorithms \cite{Cachin:2011wt}.

In the $\isa{node{\isacharunderscore}histories}$ locale we may write $\isa{x} \sqsubset^\isa{i} \isa{y}$, which means that event $\isa{x}$ \emph{comes before} event $\isa{y}$ in the history of node $\isa{i}$.
More formally, $\isa{x} \sqsubset^\isa{i} \isa{y}$ if and only if there exist lists $\isa{xs}$, $\isa{ys}$, and $\isa{zs}$ such that $\isa{xs}\mathbin{@}[\isa{x}]\mathbin{@}\isa{ys}\mathbin{@}[\isa{y}]\mathbin{@}\isa{zs} = \isa{history\ i}$.

\subsection{An asynchronous broadcast network}\label{sect.network.broadcast}

We now extend the $\isa{node-histories}$ locale by defining how nodes can communicate.
We specialise $\isacharprime\isa{a}$ to be one of two kinds of event: either \emph{broadcast} or \emph{deliver}.
(In the conventional distributed systems terminology, a \emph{deliver} event indicates that a message was received from the network and delivered to the application.)
Each event contains a message of some abstract type $\isacharprime\isa{msg}$:
\begin{isabelle}
\isacommand{datatype} {\isacharprime}msg\ event\ {\isacharequal}\ Broadcast\ {\isacharprime}msg\ {\isacharbar}\ Deliver\ {\isacharprime}msg
\end{isabelle}

Intuitively, a node can be regarded as a deterministic state machine where each state transition corresponds to a broadcast or deliver event.
We assume that users may query the state of any node at any time, and such queries need not be reflected as events, since they neither modify the node state nor send or receive any messages.

A broadcast abstraction is the standard network model for operation-based CRDTs because it best fits the replication pattern: any node can accept writes, and propagate them to the other nodes through broadcast.
In practical systems, broadcast abstractions are often implemented as overlay networks on top of unicast TCP links, for example as a fully connected graph (each node is connected to every other node), using a spanning tree protocol, a gossip protocol, or some other network topology.
Such protocols have already been studied extensively, for example by \citet{Leitao:2007gq}, so we leave the implementation of the overlay network out of the scope of this paper.

To formally specify the properties of a broadcast network, we define a new locale $\isa{network}$ containing three axioms that define how broadcast and deliver events may interact.
Since $\isa{network}$ is an extension of $\isa{node-histories}$, the aforementioned definitions of $\isa{history}$ and $\sqsubset^\isa{i}$ are available for use in the $\isa{network}$ axioms:
\begin{isabelle}
~~~~~~~~\isakeyword{and}\ \=msg{\isacharunderscore}id{\isacharunderscore}unique{\isacharcolon}\ \={\isasymrbrakk}\ \={\isachardoublequoteopen}Broadcast\ m\ {\isasymin}\ set\ {\isacharparenleft}history\ i{\isacharparenright}\ \=\kill
\isacommand{locale}\ network\ {\isacharequal}\ node{\isacharunderscore}histories\ history\\
~~~~\isakeyword{for}\>history\ {\isacharcolon}{\isacharcolon}\ {\isachardoublequoteopen}nat\ {\isasymRightarrow}\ {\isacharprime}msg\ event\ list{\isachardoublequoteclose}\ {\isacharplus}\\
~~~~\isakeyword{fixes}\>msg{\isacharunderscore}id\ {\isacharcolon}{\isacharcolon}\ {\isachardoublequoteopen}{\isacharprime}msg\ {\isasymRightarrow}\ {\isacharprime}msgid{\isachardoublequoteclose}\\
~~~~\isakeyword{assumes}\ delivery{\isacharunderscore}has{\isacharunderscore}a{\isacharunderscore}cause{\isacharcolon}\\
\>\>{\isasymlbrakk}\ {\isachardoublequoteopen}Deliver\ m\ {\isasymin}\ set\ {\isacharparenleft}history\ i{\isacharparenright}\ \>\>{\isasymrbrakk}\ {\isasymLongrightarrow}\ {\isasymexists}j{\isachardot}\ Broadcast\ m\ {\isasymin}\ set\ {\isacharparenleft}history\ j{\isacharparenright}{\isachardoublequoteclose}\\
~~~~~~~~\isakeyword{and}\>deliver{\isacharunderscore}locally{\isacharcolon}\ \>{\isasymlbrakk}\ \>{\isachardoublequoteopen}Broadcast\ m\ {\isasymin}\ set\ {\isacharparenleft}history\ i{\isacharparenright}\ \>{\isasymrbrakk}\ {\isasymLongrightarrow}\  Broadcast\ m\ {\isasymsqsubset}\isactrlsup i\ Deliver\ m{\isachardoublequoteclose}\\
~~~~~~~~\isakeyword{and}\>msg{\isacharunderscore}id{\isacharunderscore}unique{\isacharcolon}\ \>{\isasymlbrakk}\ \>{\isachardoublequoteopen}Broadcast\ m{\isadigit{1}}\ {\isasymin}\ set\ {\isacharparenleft}history\ i{\isacharparenright};\\
\>\>\>Broadcast\ m{\isadigit{2}}\ {\isasymin}\ set\ {\isacharparenleft}history\ j{\isacharparenright};\\
\>\>\>msg{\isacharunderscore}id\ m{\isadigit{1}}\ {\isacharequal}\ msg{\isacharunderscore}id\ m{\isadigit{2}}\ \>{\isasymrbrakk}\ {\isasymLongrightarrow}\ i\ {\isacharequal}\ j\ {\isasymand}\ m{\isadigit{1}}\ {\isacharequal}\ m{\isadigit{2}}{\isachardoublequoteclose}
\end{isabelle}
The axioms can be understood as follows:
\begin{description}
    \item[delivery-has-a-cause:] If some message $\isa{m}$ was delivered at some node, then there exists some node on which $\isa{m}$ was broadcast.
        With this axiom, we assert that messages are not created ``out of thin air'' by the network itself, and that the only source of messages are the nodes.
    \item[deliver-locally:] If a node broadcasts some message $\isa{m}$, then the same node must subsequently also deliver $\isa{m}$ to itself.
        Since $\isa{m}$ does not actually travel over the network, this local delivery is always possible, even if the network is interrupted.
        Local delivery may seem redundant, since its effect could also occur in the broadcast event, but it is convenient for algorithms that use the broadcast abstraction \cite{Cachin:2011wt}.
    \item[msg-id-unique:] We do not require the message type $\isacharprime\isa{msg}$ to have any particular structure; we only assume the existence of a function $\isa{msg-id} \mathbin{\isacharcolon\isacharcolon} \isacharprime\isa{msg} \mathbin{\isasymRightarrow} \isacharprime\isa{msgid}$ that maps every message to some globally unique identifier of type $\isacharprime\isa{msgid}$.
        We assert this uniqueness by stating that if $\isa{m1}$ and $\isa{m2}$ are any two messages broadcast by any two nodes, and their $\isa{msg-id}$s are the same, then they were in fact broadcast by the same node and the two messages are identical. 
        In practice, these globally unique IDs can by implemented using unique node identifiers, sequence numbers or timestamps.
\end{description}

The $\isa{network}$ locale also inherits the $\isa{histories-distinct}$ axiom from its parent locale $\isa{node-histories}$.
Many other properties that we require can be deduced as lemmas from these axioms.
For example, we can prove that for every message that is delivered by some node, there is exactly one broadcast event (on the same or some other node) that created the message.
Also, due to the $\isa{histories-distinct}$ axiom we know that the same message is not delivered more than once to each node---an aspect that can be implemented in practical systems by having each node keep track of message IDs it has received, and suppressing any duplicates.

Note that we make no assumptions about the reliability or the ordering of messages.
If one node broadcasts a message, it \emph{may} be delivered by other nodes, but we do not state if or when that will happen.
Messages may be arbitrarily delayed, reordered, or even lost entirely.
It is even acceptable for a node to never deliver any messages besides those it broadcasts itself, modelling a node that is permanently disconnected from the network.

\subsection{Causally ordered delivery}\label{sect.network.causal}

As discussed in Section~\ref{sect.happens.before}, some replication algorithms require that some operations be applied in a particular order because the later operation has a causal dependency on the earlier one.
We previously characterised these dependencies using the \emph{happens-before} relation $\prec$, which we required to be a strict partial order, but otherwise kept abstract.
In Section~\ref{sect.abstract.convergence} we reasoned about the order of \emph{operations}, but in a network we work with \emph{messages}.
We will connect operations and messages in Section~\ref{sect.network.ops}; for now we will define a particular instance of the ordering relation $\prec$ on messages, and prove that it satisfies the requirements of a strict partial order.

We do not use physical time (such as UTC) to define the order of messages, since reliance on physical time is often problematic in distributed systems \cite{Sheehy:2015jm}.
Instead, we say that a message $\isa{m1}$ happens before another message $\isa{m2}$ if the node that generated $\isa{m2}$ ``knew about'' $\isa{m1}$ at the time $\isa{m2}$ was generated.
More precisely, based on the well-known definition by \citet{Lamport:1978jq}, we say that $\isa{m1}\prec\isa{m2}$ if any of the following is true:
\begin{enumerate}
\item $\isa{m1}$ and $\isa{m2}$ were broadcast by the same node, and $\isa{m1}$ was broadcast before $\isa{m2}$.
\item The node that broadcast $\isa{m2}$ had delivered $\isa{m1}$ before it broadcast $\isa{m2}$.
\item There exists some operation $\isa{m3}$ such that $\isa{m1} \prec \isa{m3}$ and $\isa{m3} \prec \isa{m2}$.
\end{enumerate}

This verbal definition translates directly into Isabelle syntax:
\begin{isabelle}
~~~~{\isachardoublequoteopen}{\isasymlbrakk}\ Broadcast\ m{\isadigit{1}}\ \={\isasymsqsubset}\isactrlsup i\ Broadcast\ m{\isadigit{2}}\ \=\kill
\isacommand{inductive}\ hb\ {\isacharcolon}{\isacharcolon}\ {\isachardoublequoteopen}{\isacharprime}msg\ {\isasymRightarrow}\ {\isacharprime}msg\ {\isasymRightarrow}\ bool{\isachardoublequoteclose}\ \isakeyword{where}\\
~~~~{\isachardoublequoteopen}{\isasymlbrakk}\ Broadcast\ m{\isadigit{1}}\ \>{\isasymsqsubset}\isactrlsup i\ Broadcast\ m{\isadigit{2}}\ \>{\isasymrbrakk}\ {\isasymLongrightarrow}\ m{\isadigit{1}}\ $\prec$\ m{\isadigit{2}}{\isachardoublequoteclose}\ {\isacharbar}\\
~~~~{\isachardoublequoteopen}{\isasymlbrakk}\ Deliver\ m{\isadigit{1}}\ \>{\isasymsqsubset}\isactrlsup i\ Broadcast\ m{\isadigit{2}}\ \>{\isasymrbrakk}\ {\isasymLongrightarrow}\ m{\isadigit{1}}\ $\prec$\ m{\isadigit{2}}{\isachardoublequoteclose}\ {\isacharbar}\\
~~~~{\isachardoublequoteopen}{\isasymlbrakk}\ m{\isadigit{1}}\ $\prec$\  m{\isadigit{2}}{\isacharsemicolon}\ \ m{\isadigit{2}}\ $\prec$\ m{\isadigit{3}}\ \>\>{\isasymrbrakk}\ {\isasymLongrightarrow}\ m{\isadigit{1}}\ $\prec$\ m{\isadigit{3}}{\isachardoublequoteclose}
\end{isabelle}
Given this definition, we define a restricted variant of our broadcast network model by extending the $\isa{network}$ locale.
In addition to the existing $\isa{network}$ axioms, we require that if there are any happens-before dependencies between messages, they must be delivered in that order.
Concurrent messages may be delivered in any order.
\begin{isabelle}
\isacommand{locale} causal{\isacharunderscore}network\ {\isacharequal}\ network\ {\isacharplus}\\
~~~~\isakeyword{assumes}\ causal{\isacharunderscore}delivery{\isacharcolon}\\
~~~~~~~~{\isasymlbrakk}\ {\isachardoublequoteopen}Deliver\ m{\isadigit{2}}\ {\isasymin}\ set\ {\isacharparenleft}history\ i{\isacharparenright};\ m{\isadigit{1}}\ $\prec$\ m{\isadigit{2}}\ {\isasymrbrakk}\ {\isasymLongrightarrow}\ Deliver\ m{\isadigit{1}}\ {\isasymsqsubset}\isactrlsup i\ Deliver\ m{\isadigit{2}}{\isachardoublequoteclose}
\end{isabelle}
The $\isa{causal-delivery}$ axiom does not strengthen the reliability assumptions of the network: only in the case where some message $\isa{m2}$ is delivered, it requires that any causally preceding messages are delivered first.
It is still possible for some message never to be delivered.
Causal delivery is typically implemented in network protocols using vector timestamps \cite{Schwarz:1994gl,Fidge:1988tv,Raynal:1996jl}.
As these protocols are widely known and well understood, we elide any further discussion.

\subsection{Using operations in the network}\label{sect.network.ops}

We can now include the convergence theorem into our network model by further extending the $\isa{causal-network}$ locale.
In the new locale $\isa{network-with-ops}$ we do not assume any additional axioms; we only specialise the type variable of messages $\isacharprime\isa{msg}$ to be a pair of $\isacharprime\isa{msgid} \mathbin{\isasymtimes} \isacharprime\isa{oper}$, and we instantiate the $\isa{msg-id}$ function fixed by the $\isa{network}$ locale to be $\isa{fst}$, i.e., to return the first component $\isacharprime\isa{msgid}$ of the pair.
We also assume the existence of an interpretation function (see Section~\ref{sect.ops.interpretation}) and a fixed initial node state:
\begin{isabelle}
~~~~\isakeyword{fixes}\ \=history\ \=\kill
\isacommand{locale}\ network{\isacharunderscore}with{\isacharunderscore}ops\ {\isacharequal}\ causal{\isacharunderscore}network\ history\ fst\\
~~~~\isakeyword{for}\ \>history\ \>{\isacharcolon}{\isacharcolon}\ {\isachardoublequoteopen}nat\ {\isasymRightarrow}\ {\isacharparenleft}{\isacharprime}msgid\ {\isasymtimes}\ {\isacharprime}oper{\isacharparenright}\ event\ list{\isachardoublequoteclose}\ {\isacharplus}\\
~~~~\isakeyword{fixes}\ \>interp\ \>{\isacharcolon}{\isacharcolon}\ {\isachardoublequoteopen}{\isacharprime}oper\ {\isasymRightarrow}\ {\isacharprime}state\ {\isasymRightarrow}\ {\isacharprime}state\ option{\isachardoublequoteclose}\\
~~~~\isakeyword{and}\ \>initial{\isacharunderscore}state\ {\isacharcolon}{\isacharcolon}\ {\isachardoublequoteopen}{\isacharprime}state{\isachardoublequoteclose}
\end{isabelle}
We have proved that the happens-before relation $\prec$ defined in the network is a strict partial order, so it meets the requirements of the $\isa{happens-before}$ locale.
The lemmas and definitions of this locale are therefore available to use with the happens-before relation $\prec$, and we indicate these specialised theorems and definitions by prefixing their names with $\isa{hb}$.
Moreover, we can prove that the sequence of message deliveries at any node is consistent with $\prec$, that is, it satisfies the definition of $\isa{hb-consistent}$ given in Section~\ref{sect.happens.before} (note $\isa{hb}{\isacharunderscore}\isa{consistent}$ is now prefixed):
\begin{isabelle}
\isacommand{theorem}\ \ {\isachardoublequoteopen}hb.hb{\isacharunderscore}consistent\ {\isacharparenleft}node{\isacharunderscore}deliver{\isacharunderscore}messages\ {\isacharparenleft}history\ i{\isacharparenright}{\isacharparenright}{\isachardoublequoteclose}
\end{isabelle}
\noindent
where $\isa{node-deliver-messages}$ is a function that filters the history of events at some node to return only messages that were delivered, in the order they were delivered.
Now, whenever a message is delivered at some node, we can take the operation $\isacharprime\isa{oper}$ from the message, and use its interpretation to update the state at that node.
Broadcast events do not change the state, but since every message must be delivered locally at the node where it was broadcast, the state change nevertheless takes effect locally.
We can then define the state of some node by using our definition of $\isa{apply-operations}$ from Section~\ref{sect.ops.interpretation}:
\begin{isabelle}
\isacommand{definition}\ apply{\isacharunderscore}operations\ {\isacharcolon}{\isacharcolon}\ {\isachardoublequoteopen}{\isacharparenleft}{\isacharprime}msgid\ {\isasymtimes}\ {\isacharprime}oper{\isacharparenright}\ event\ list\ {\isasymRightarrow}\ {\isacharprime}state\ option{\isachardoublequoteclose}\ \isakeyword{where}\\
~~~~{\isachardoublequoteopen}apply{\isacharunderscore}operations\ es\ {\isasymequiv}\ hb{\isachardot}apply{\isacharunderscore}operations\ {\isacharparenleft}node{\isacharunderscore}deliver{\isacharunderscore}messages\ es{\isacharparenright}\ initial{\isacharunderscore}state{\isachardoublequoteclose}
\end{isabelle}

So far we have no restriction on the operations that may be broadcast, except that they must be of some type $\isacharprime\isa{oper}$.
This suffices for some replication algorithms, but many have additional requirements regarding the contents of messages that cannot be expressed in Isabelle's type system.
As a general-purpose means of describing such requirements, the locale $\isa{network-with-constrained-ops}$ allows a replication algorithm to define a predicate $\isa{valid-msg}$ to specify whether a node is allowed to broadcast some message when in a particular state:
\begin{isabelle}
\hspace{6em}\=\kill
\isacommand{locale}\ network{\isacharunderscore}with{\isacharunderscore}constrained{\isacharunderscore}ops\ {\isacharequal}\ network{\isacharunderscore}with{\isacharunderscore}ops\ {\isacharplus}\\
~~~~\isakeyword{fixes}\ valid{\isacharunderscore}msg\ {\isacharcolon}{\isacharcolon}\ {\isachardoublequoteopen}{\isacharprime}state\ {\isasymRightarrow}\ {\isacharparenleft}{\isacharprime}msgid\ {\isasymtimes}\ {\isacharprime}oper{\isacharparenright}\ {\isasymRightarrow}\ bool{\isachardoublequoteclose}\\
~~~~\isakeyword{assumes}\ broadcast{\isacharunderscore}only{\isacharunderscore}valid{\isacharunderscore}msgs{\isacharcolon}\\
\>{\isachardoublequoteopen}{\isasymexists}suf{\isachardot}\ pre\ {\isacharat}\ {\isacharbrackleft}Broadcast\ m{\isacharbrackright}\ {\isacharat}\ suf\ {\isacharequal}\ history\ i\ {\isasymLongrightarrow}\\
\>{\isasymexists}state{\isachardot}\ apply{\isacharunderscore}operations\ pre\ {\isacharequal}\ Some\ state\ {\isasymand}\ valid{\isacharunderscore}msg\ state\ m{\isachardoublequoteclose}
\end{isabelle}

$\isa{broadcast-only-valid-msgs}$ is our final axiom, and it simply requires that if a node broadcasts some message, it must be valid according to the $\isa{valid-msg}$ predicate.
Since the choice of messages to broadcast is under the control of the replication algorithm, and the algorithm defines this predicate, this assumption is reasonable.

Although these six axioms are simple and uncontroversial, we believe that the set of axioms could be reduced further by defining some of the aforementioned algorithms (such as vector timestamps for causal delivery, or sequence numbers for message uniqueness) within Isabelle, and proving that the algorithms guarantee the required properties within some weaker network model.
However, doing so would lead us too far astray from the goal of proving the strong eventual consistency of CRDTs, so we leave it for future work.

The axioms of $\isa{network}{\isacharunderscore}\isa{with}{\isacharunderscore}\isa{constrained}{\isacharunderscore}\isa{ops}$ and its superlocales are consistent (in the sense that that we are unable to prove $\isa{False}$ by assuming the axioms).
We demonstrate this fact by building a trivial model of $\isa{network}{\isacharunderscore}\isa{with}{\isacharunderscore}\isa{constrained}{\isacharunderscore}\isa{ops}$ within Isabelle and showing that it satisfies all of the locale's axioms.
We elide these models here.

\section{Replicated Growable Array}
\label{sect.rga}

The RGA, introduced by \citet{Roh:2011dw}, is a replicated ordered list (sequence) datatype that supports \emph{insert} and \emph{delete} operations.
It can be used for collaborative editing of text by representing a string as an ordered list of characters.

The convergence of RGA has been proved by hand in previous work (see Section~\ref{sect.related.verification}); we now present the first (to our knowledge) mechanised proof that RGA satisfies the specification of SEC from Section~\ref{sect.abstract.convergence}.
We perform this proof within the causal broadcast model defined in Section~\ref{sect.network}, and without making any assumptions beyond the six aforementioned network axioms.
Since the axioms of our network model are easily justified, we have confidence in the correctness of our formalisation.
Our proof makes extensive use of the general-purpose framework that we have established in the last two sections.

\subsection{Specifying insertion and deletion}\label{sect.rga.spec}

In an ordered list, each insertion and deletion operation must identify the position at which the modification should take place.
In a non-replicated setting, the position is commonly expressed as an index into the list.
However, the index of a list element may change if other elements are concurrently inserted or deleted earlier in the list; this is the problem at the heart of Operational Transformation (see Section~\ref{sect.related.ot.crdts}).
Instead of using indexes, the RGA algorithm assigns a unique, immutable identifier to each list element.

Insertion operations place the new element \emph{after} an existing list element with a given ID, or at the head of the list if no ID is given.
Deletion operations refer to the ID of the list element that is to be deleted.
However, it is not safe for a deletion operation to completely remove a list element, because then a concurrent insertion after the deleted element would not be able to locate the insertion position.
Instead, the list retains \emph{tombstones}: a deletion operation merely sets a flag on a list element to mark it as deleted, but the element actually remains in the list.
A garbage collection process can be used to purge tombstones \cite{Roh:2011dw}, but we do not consider it here.

The RGA state at each node is a list of elements.
Each element is a triple consisting of the unique ID of the list element (of some type $\isacharprime\isa{id}$), the value inserted by the application (of some type $\isacharprime\isa{v}$), and a flag that indicates whether the element has been marked as deleted (of type $\isa{bool}$):
\begin{isabelle}
\isacommand{type{\isacharunderscore}synonym} {\isacharparenleft}{\isacharprime}id{\isacharcomma}\ {\isacharprime}v{\isacharparenright}\ elt\ {\isacharequal}\ {\isachardoublequoteopen}{\isacharprime}id\ {\isasymtimes}\ {\isacharprime}v\ {\isasymtimes}\ bool{\isachardoublequoteclose}%
\end{isabelle}

The $\isa{insert}$ function takes three parameters: the previous state of the list, the new element to insert, and optionally the ID of an existing element after which the new element should be inserted.
It returns the list with the new element inserted at the appropriate position, or $\isa{None}$ on failure, which occurs if there was no existing element with the given ID.
The function iterates over the list, and for each list element $\isa{x}$, it compares the ID (the first component of the $\isacharprime\isa{id} \mathbin{\isasymtimes} \isacharprime\isa{v} \mathbin{\isasymtimes} \isa{bool}$ triple, written $\isa{fst x}$) to the requested insertion position:
\begin{isabelle}
~~~~{\isachardoublequoteopen}insert\ {\isacharparenleft}x{\isacharhash}xs{\isacharparenright}\ \=e\ {\isacharparenleft}Some\ i{\isacharparenright}\ \={\isacharequal}\ {\isacharparenleft}\=\kill
\isacommand{fun}\ insert\ {\isacharcolon}{\isacharcolon}\ {\isachardoublequoteopen}{\isacharparenleft}{\isacharprime}id{\isacharcolon}{\isacharcolon}{\isacharbraceleft}linorder{\isacharbraceright}{\isacharcomma}\ {\isacharprime}v{\isacharparenright}\ elt\ list\ {\isasymRightarrow}\ {\isacharparenleft}{\isacharprime}id{\isacharcomma}\ {\isacharprime}v{\isacharparenright}\ elt\ {\isasymRightarrow}\ {\isacharprime}id\ option\ {\isasymRightarrow}\ {\isacharparenleft}{\isacharprime}id{\isacharcomma}\ {\isacharprime}v{\isacharparenright}\ elt\ list\ option{\isachardoublequoteclose}\\
\isakeyword{where}\\
~~~~{\isachardoublequoteopen}insert\ xs\ \>e\ None\ \>{\isacharequal}\ Some\ {\isacharparenleft}insert{\isacharunderscore}body\ xs\ e{\isacharparenright}{\isachardoublequoteclose}\ {\isacharbar}\\
~~~~{\isachardoublequoteopen}insert\ {\isacharbrackleft}{\isacharbrackright}\ \>e\ {\isacharparenleft}Some\ i{\isacharparenright}\ \>{\isacharequal}\ None{\isachardoublequoteclose}\ {\isacharbar}\\
~~~~{\isachardoublequoteopen}insert\ {\isacharparenleft}x{\isacharhash}xs{\isacharparenright}\ \>e\ {\isacharparenleft}Some\ i{\isacharparenright}\ \>{\isacharequal}\ {\isacharparenleft}\>if\ fst\ x\ {\isacharequal}\ i\ then\ Some\ {\isacharparenleft}x{\isacharhash}insert{\isacharunderscore}body\ xs\ e{\isacharparenright}\\
\>\>\>else\ insert\ xs\ e\ {\isacharparenleft}Some\ i{\isacharparenright}\ {\isasymbind}\ {\isacharparenleft}{\isasymlambda}t{\isachardot}\ Some\ {\isacharparenleft}x{\isacharhash}t{\isacharparenright}{\isacharparenright}{\isacharparenright}{\isachardoublequoteclose}
\end{isabelle}
When the insertion position is found (or, in the case of insertion at the head of the list, immediately), the function $\isa{insert-body}$ is invoked to perform the actual insertion:
\begin{isabelle}
~~~~{\isachardoublequoteopen}insert{\isacharunderscore}body\ {\isacharparenleft}x{\isacharhash}xs{\isacharparenright}\ \=\kill
\isacommand{fun} insert{\isacharunderscore}body\ {\isacharcolon}{\isacharcolon}\ {\isachardoublequoteopen}{\isacharparenleft}{\isacharprime}id{\isacharcolon}{\isacharcolon}{\isacharbraceleft}linorder{\isacharbraceright}{\isacharcomma}\ {\isacharprime}v{\isacharparenright}\ elt\ list\ {\isasymRightarrow}\ {\isacharparenleft}{\isacharprime}id{\isacharcomma}\ {\isacharprime}v{\isacharparenright}\ elt\ {\isasymRightarrow}\ {\isacharparenleft}{\isacharprime}id{\isacharcomma}\ {\isacharprime}v{\isacharparenright}\ elt\ list{\isachardoublequoteclose}\ \isakeyword{where}\\
~~~~{\isachardoublequoteopen}insert{\isacharunderscore}body\ {\isacharbrackleft}{\isacharbrackright} \>e\ {\isacharequal}\ {\isacharbrackleft}e{\isacharbrackright}{\isachardoublequoteclose}\ {\isacharbar}\\
~~~~{\isachardoublequoteopen}insert{\isacharunderscore}body\ {\isacharparenleft}x{\isacharhash}xs{\isacharparenright}\>e\ {\isacharequal}\ {\isacharparenleft}if\ fst\ x\ {\isacharless}\ fst\ e\ then\ e{\isacharhash}x{\isacharhash}xs\ else\ x{\isacharhash}insert{\isacharunderscore}body\ xs\ e{\isacharparenright}{\isachardoublequoteclose}
\end{isabelle}

In a non-replicated datatype it would be sufficient to insert the new element directly at the position found by the $\isa{insert}$ function.
However, a replicated setting is more difficult, because several nodes may concurrently insert new elements at the same position, and those insertion operations may be processed in a different order by different nodes.
In order to ensure that all nodes converge towards the same state (that is, the same order of list elements), we sort any concurrent insertions at the same position in descending order of the inserted elements' IDs.
This sorting is implemented in $\isa{insert-body}$ by skipping over any elements with an ID that is greater than that of the newly inserted element (the $\isa{fst x} > \isa{fst e}$ case), and then placing the new element before the first existing element with a lesser ID (the $\isa{fst x} < \isa{fst e}$ case).

Note that the type of IDs is specified as $\isacharprime\isa{id}\isacharcolon\isacharcolon\isacharbraceleft\isa{linorder}\isacharbraceright$, which means that we require the type $\isacharprime\isa{id}$ to have an associated total (linear) order.
$\isa{linorder}$ is the name of a type class supplied by the Isabelle/HOL library.
This annotation is required in order to be able to perform the comparison $\isa{fst x} < \isa{fst e}$ on IDs.
To be precise, RGA requires the total order of IDs to be consistent with causality, which can easily be achieved using the logical timestamps defined by \citet{Lamport:1978jq}.

The delete operation searches for the element with a given ID, and sets its flag to $\isa{True}$ to mark it as deleted:
\begin{isabelle}
~~~~{\isachardoublequoteopen}delete\ {\isacharparenleft}{\isacharparenleft}i{\isacharprime}{\isacharcomma}\ v{\isacharcomma}\ flag{\isacharparenright}{\isacharhash}xs{\isacharparenright}\ \=i\ {\isacharequal}\ {\isacharparenleft}\=\kill
\isacommand{fun}\ delete\ {\isacharcolon}{\isacharcolon}\ {\isachardoublequoteopen}{\isacharparenleft}{\isacharprime}id{\isacharcolon}{\isacharcolon}{\isacharbraceleft}linorder{\isacharbraceright}{\isacharcomma}\ {\isacharprime}v{\isacharparenright}\ elt\ list\ {\isasymRightarrow}\ {\isacharprime}id\ {\isasymRightarrow}\ {\isacharparenleft}{\isacharprime}id{\isacharcomma}\ {\isacharprime}v{\isacharparenright}\ elt\ list\ option{\isachardoublequoteclose}\ \isakeyword{where}\\
~~~~{\isachardoublequoteopen}delete\ {\isacharbrackleft}{\isacharbrackright}\>i\ {\isacharequal}\ None{\isachardoublequoteclose}\ {\isacharbar}\\
~~~~{\isachardoublequoteopen}delete\ {\isacharparenleft}{\isacharparenleft}i{\isacharprime}{\isacharcomma}\ v{\isacharcomma}\ flag{\isacharparenright}{\isacharhash}xs{\isacharparenright} \>i\ {\isacharequal}\ {\isacharparenleft}\>if\ i{\isacharprime}\ {\isacharequal}\ i\ then\ Some\ {\isacharparenleft}{\isacharparenleft}i{\isacharprime}{\isacharcomma}\ v{\isacharcomma}\ True{\isacharparenright}{\isacharhash}xs{\isacharparenright}\\
\>\>else\ delete\ xs\ i\ {\isasymbind}\ {\isacharparenleft}{\isasymlambda}t{\isachardot}\ Some\ {\isacharparenleft}{\isacharparenleft}i{\isacharprime}{\isacharcomma}v{\isacharcomma}flag{\isacharparenright}{\isacharhash}t{\isacharparenright}{\isacharparenright}{\isacharparenright}{\isachardoublequoteclose}%
\end{isabelle}
Note that the operations presented here are deliberately inefficient in order to make them easier to reason about.
One can see our implementations of $\isa{insert-body}$, $\isa{insert}$, and $\isa{delete}$ as functional specifications for RGAs, which could be optimised into more efficient algorithms using data refinement, if desired.

\subsection{Commutativity of insertion and deletion}

Recall from Section~\ref{sect.ops.commute} that in order to prove the convergence theorem we need to show that for the datatype in question, all its concurrent operations commute.
It is straightforward to demonstrate that $\isa{delete}$ always commutes with itself, on concurrent and non-concurrent operations alike:
\begin{isabelle}
\isacommand{lemma}\ delete{\isacharunderscore}commutes{\isacharcolon}\\
~~~~{\isachardoublequoteopen}delete\ xs\ i{\isadigit{1}}\ {\isasymbind}\ {\isacharparenleft}{\isasymlambda}ys{\isachardot}\ delete\ ys\ i{\isadigit{2}}{\isacharparenright}\ {\isacharequal}\ delete\ xs\ i{\isadigit{2}}\ {\isasymbind}\ {\isacharparenleft}{\isasymlambda}ys{\isachardot}\ delete\ ys\ i{\isadigit{1}}{\isacharparenright}{\isachardoublequoteclose}
\end{isabelle}

It is a little more complex to demonstrate that two $\isa{insert}$ operations commute.
Let $\isa{e1}$ and $\isa{e2}$ be the two new list elements being inserted, each of which is a $\isacharprime\isa{id} \mathbin{\isasymtimes} \isacharprime\isa{v} \mathbin{\isasymtimes} \isa{bool}$ triple.
Further, let $\isa{i1} \mathbin{\isacharcolon\isacharcolon} \isacharprime\isa{id option}$ be the position after which $\isa{e1}$ should be inserted (either $\isa{None}$ for the head of the list, or $\isa{Some i}$ where $\isa{i}$ is the ID of an existing list element), and similarly let $\isa{i2}$ be the position after which $\isa{e2}$ should be inserted.
Then the two insertions commute only under certain assumptions:
\begin{isabelle}
~~~~\isakeyword{assumes}\ \=\kill
\isacommand{lemma}\ insert{\isacharunderscore}commutes{\isacharcolon}\\
~~~~\isakeyword{assumes}\>{\isachardoublequoteopen}fst\ e{\isadigit{1}}\ {\isasymnoteq}\ fst\ e{\isadigit{2}}{\isachardoublequoteclose}\\
~~~~~~~~\isakeyword{and}\>{\isachardoublequoteopen}i{\isadigit{1}}\ {\isacharequal}\ None\ {\isasymor}\ i{\isadigit{1}}\ {\isasymnoteq}\ Some\ {\isacharparenleft}fst\ e{\isadigit{2}}{\isacharparenright}{\isachardoublequoteclose}\\
~~~~~~~~\isakeyword{and}\>{\isachardoublequoteopen}i{\isadigit{2}}\ {\isacharequal}\ None\ {\isasymor}\ i{\isadigit{2}}\ {\isasymnoteq}\ Some\ {\isacharparenleft}fst\ e{\isadigit{1}}{\isacharparenright}{\isachardoublequoteclose}\\
~~~~\isakeyword{shows}\>{\isachardoublequoteopen}insert\ xs\ e{\isadigit{1}}\ i{\isadigit{1}}\ {\isasymbind}\ {\isacharparenleft}{\isasymlambda}ys{\isachardot}\ insert\ ys\ e{\isadigit{2}}\ i{\isadigit{2}}{\isacharparenright}\ {\isacharequal}\ insert\ xs\ e{\isadigit{2}}\ i{\isadigit{2}}\ {\isasymbind}\ {\isacharparenleft}{\isasymlambda}ys{\isachardot}\ insert\ ys\ e{\isadigit{1}}\ i{\isadigit{1}}{\isacharparenright}{\isachardoublequoteclose}
\end{isabelle}
\noindent
That is, $\isa{i1}$ cannot refer to the ID of $\isa{e2}$ and vice versa, and the IDs of the two insertions must be distinct.
We prove later that these assumptions are indeed satisfied for all concurrent operations.
Finally, $\isa{delete}$ commutes with $\isa{insert}$ whenever the element to be deleted is not the same as the element to be inserted:
\begin{isabelle}
~~~~\isakeyword{assumes}\ \=\kill
\isacommand{lemma}\ insert{\isacharunderscore}delete{\isacharunderscore}commute{\isacharcolon}\\
~~~~\isakeyword{assumes}\>{\isachardoublequoteopen}i{\isadigit{2}}\ {\isasymnoteq}\ fst\ e{\isachardoublequoteclose}\\
~~~~\isakeyword{shows}\>{\isachardoublequoteopen}insert\ xs\ e\ i{\isadigit{1}}\ {\isasymbind}\ {\isacharparenleft}{\isasymlambda}ys{\isachardot}\ delete\ ys\ i{\isadigit{2}}{\isacharparenright}\ {\isacharequal}\ delete\ xs\ i{\isadigit{2}}\ {\isasymbind}\ {\isacharparenleft}{\isasymlambda}ys{\isachardot}\ insert\ ys\ e\ i{\isadigit{1}}{\isacharparenright}{\isachardoublequoteclose}
\end{isabelle}

\subsection{Embedding RGA in the network model}

In order to obtain a proof of the strong eventual consistency of RGA, we embed the insertion and deletion operations in the network model of Section~\ref{sect.network}.
We first define a datatype for operations (which are sent across the network in messages), and an interpretation function as introduced in Section~\ref{sect.ops.interpretation}:
\begin{isabelle}
~~~~{\isachardoublequoteopen}interpret{\isacharunderscore}opers\ {\isacharparenleft}Insert\ e\ n{\isacharparenright}\ \=\kill
\isacommand{datatype} {\isacharparenleft}{\isacharprime}id{\isacharcomma}\ {\isacharprime}v{\isacharparenright}\ operation\ {\isacharequal} Insert\ {\isachardoublequoteopen}{\isacharparenleft}{\isacharprime}id{\isacharcomma}\ {\isacharprime}v{\isacharparenright}\ elt{\isachardoublequoteclose}\ {\isachardoublequoteopen}{\isacharprime}id\ option{\isachardoublequoteclose}\ {\isacharbar} Delete\ {\isachardoublequoteopen}{\isacharprime}id{\isachardoublequoteclose}\\[4pt]
\isacommand{fun} interpret{\isacharunderscore}opers\ {\isacharcolon}{\isacharcolon}\ {\isachardoublequoteopen}{\isacharparenleft}{\isacharprime}id{\isacharcolon}{\isacharcolon}linorder{\isacharcomma}\ {\isacharprime}v{\isacharparenright}\ operation\ {\isasymRightarrow}\ {\isacharparenleft}{\isacharprime}id{\isacharcomma}\ {\isacharprime}v{\isacharparenright}\ elt\ list\ {\isasymRightarrow}\ {\isacharparenleft}{\isacharprime}id{\isacharcomma}\ {\isacharprime}v{\isacharparenright}\ elt\ list\ option{\isachardoublequoteclose}\\
\isakeyword{where}\\
~~~~{\isachardoublequoteopen}interpret{\isacharunderscore}opers\ {\isacharparenleft}Insert\ e\ n{\isacharparenright}\>xs\ \ {\isacharequal}\ insert\ xs\ e\ n{\isachardoublequoteclose}\ {\isacharbar}\\
~~~~{\isachardoublequoteopen}interpret{\isacharunderscore}opers\ {\isacharparenleft}Delete\ n{\isacharparenright}\>xs\ \ {\isacharequal}\ delete\ xs\ n{\isachardoublequoteclose}
\end{isabelle}

As discussed above, the validity of operations depends on some assumptions: IDs of insertion operations must be unique, and whenever an insertion or deletion operation refers to an existing list element, that element must exist.
As introduced in Section~\ref{sect.network.ops}, we can describe these requirements by using a predicate to specify what messages a node is allowed to broadcast when in a particular state:
\begin{isabelle}
~~~~~~~~\={\isacharparenleft}i{\isacharcomma}\ Insert\ e\ {\isacharparenleft}\=Some\ \=pos{\isacharparenright}\={\isacharparenright}\ \={\isasymRightarrow}\ fst\ e\ {\isacharequal}\ i\ {\isasymand}\ \=\kill
\isacommand{definition}\ valid{\isacharunderscore}rga{\isacharunderscore}msg\ {\isacharcolon}{\isacharcolon}\ {\isachardoublequoteopen}{\isacharparenleft}{\isacharprime}id{\isacharcomma}\ {\isacharprime}v{\isacharparenright}\ elt\ list\ {\isasymRightarrow}\ {\isacharprime}id\ {\isasymtimes}\ {\isacharparenleft}{\isacharprime}id{\isacharcolon}{\isacharcolon}linorder{\isacharcomma}\ {\isacharprime}v{\isacharparenright}\ operation\ {\isasymRightarrow}\ bool{\isachardoublequoteclose}\ \isakeyword{where}\\
~~~~{\isachardoublequoteopen}valid{\isacharunderscore}rga{\isacharunderscore}msg\ list\ msg\ {\isasymequiv}\ case\ msg\ of\\
\>{\isacharparenleft}i{\isacharcomma}\ Insert\ e \>None \>\>{\isacharparenright}\ \>{\isasymRightarrow}\ fst\ e\ {\isacharequal}\ i\ {\isacharbar}\\
\>{\isacharparenleft}i{\isacharcomma}\ Insert\ e\ {\isacharparenleft}Some\ pos{\isacharparenright}{\isacharparenright}\ {\isasymRightarrow}\ fst\ e\ {\isacharequal}\ i\ {\isasymand}\ pos\ {\isasymin}\ set\ {\isacharparenleft}map\ fst\ list{\isacharparenright}\ {\isacharbar}\\
\>{\isacharparenleft}i{\isacharcomma}\ Delete \>\>pos \>{\isacharparenright} \>{\isasymRightarrow} \>pos\ {\isasymin}\ set\ {\isacharparenleft}map\ fst\ list{\isacharparenright}{\isachardoublequoteclose}
\end{isabelle}
We can now define RGA by extending $\isa{network-with-constrained-ops}$. The interpretation function is instantiated with $\isa{interpret-opers}$, the initial state with the empty list $\isacharbrackleft\,\isacharbrackright$, and the validity predicate with $\isa{valid-rga-msg}$:
\begin{isabelle}
\isacommand{locale}\ rga\ {\isacharequal}\ network{\isacharunderscore}with{\isacharunderscore}constrained{\isacharunderscore}ops\ {\isacharunderscore}\ interpret{\isacharunderscore}opers\ {\isachardoublequoteopen}{\isacharbrackleft}{\isacharbrackright}{\isachardoublequoteclose}\ valid{\isacharunderscore}rga{\isacharunderscore}msg
\end{isabelle}

Within this locale, we prove that whenever an insertion or deletion operation $\isa{op2}$ references an existing list element, there is always a prior insertion operation $\isa{op1}$ that created the element being referenced:
\begin{isabelle}
~~~~\isakeyword{assumes}\ \={\isachardoublequoteopen}n\ {\isacharequal}\ None\ {\isasymor}\ {\isacharparenleft}{\isasymexists}e{\isacharprime}\ n{\isacharprime}{\isachardot}\ \=\kill
\isacommand{lemma}\ allowed{\isacharunderscore}insert{\isacharcolon}\\
~~~~\isakeyword{assumes}\>{\isachardoublequoteopen}Broadcast\ {\isacharparenleft}Insert\ e\ n{\isacharparenright}\ {\isasymin}\ set\ {\isacharparenleft}history\ i{\isacharparenright}{\isachardoublequoteclose}\\
~~~~\isakeyword{shows}\>{\isachardoublequoteopen}n\ {\isacharequal}\ None\ {\isasymor}\ {\isacharparenleft}{\isasymexists}e{\isacharprime}\ n{\isacharprime}{\isachardot}\>n\ {\isacharequal}\ Some\ {\isacharparenleft}fst\ e{\isacharprime}{\isacharparenright}\ {\isasymand}\\
\>\>Deliver\ {\isacharparenleft}Insert\ e{\isacharprime}\ n{\isacharprime}{\isacharparenright}\ {\isasymsqsubset}\isactrlsup i\ Broadcast\ {\isacharparenleft}Insert\ e\ n{\isacharparenright}{\isacharparenright}{\isachardoublequoteclose}\\[4pt]
\isacommand{lemma}\ allowed{\isacharunderscore}delete{\isacharcolon}\\
~~~~\isakeyword{assumes}\>{\isachardoublequoteopen}Broadcast\ {\isacharparenleft}Delete\ x{\isacharparenright}\ {\isasymin}\ set\ {\isacharparenleft}history\ i{\isacharparenright}{\isachardoublequoteclose}\\
~~~~\isakeyword{shows}\>{\isachardoublequoteopen}{\isasymexists}n{\isacharprime}\ v\ b{\isachardot}\ Deliver\ {\isacharparenleft}Insert\ {\isacharparenleft}x{\isacharcomma}\ v{\isacharcomma}\ b{\isacharparenright}\ n{\isacharprime}{\isacharparenright}\ {\isasymsqsubset}\isactrlsup i\ Broadcast\ {\isacharparenleft}Delete\ x{\isacharparenright}{\isachardoublequoteclose}
\end{isabelle}
Since the network ensures causally ordered delivery, all nodes must deliver the insertion $\isa{op1}$ before the dependent operation $\isa{op2}$.
Hence we show that in all cases where operations do not commute, one operation happens before another.
Conversely, whenever operations are concurrent, we show that they commute:
\begin{isabelle}
\isacommand{theorem}\ concurrent{\isacharunderscore}operations{\isacharunderscore}commute{\isacharcolon}\\
~~~~\isakeyword{shows}\ {\isachardoublequoteopen}hb{\isachardot}concurrent{\isacharunderscore}ops{\isacharunderscore}commute\ {\isacharparenleft}node{\isacharunderscore}deliver{\isacharunderscore}messages\ {\isacharparenleft}history\ i{\isacharparenright}{\isacharparenright}{\isachardoublequoteclose}
\end{isabelle}
Furthermore, although the type signature of the interpretation function allows an operation to fail by returning $\isa{None}$, we can prove that this failure case is never reached in any execution of the network:
\begin{isabelle}
\isacommand{theorem}\ apply{\isacharunderscore}operations{\isacharunderscore}never{\isacharunderscore}fails{\isacharcolon}\\
~~~~\isakeyword{shows}\ {\isachardoublequoteopen}hb.apply{\isacharunderscore}operations\ {\isacharparenleft}node{\isacharunderscore}deliver{\isacharunderscore}messages\ {\isacharparenleft}history\ i{\isacharparenright}{\isacharparenright}\ {\isasymnoteq}\ None{\isachardoublequoteclose}
\end{isabelle}
It is now easy to show that the $\isa{rga}$ locale satisfies all of the requirements of the abstract specification $\isa{strong-eventual-consistency}$ (Section~\ref{sect.abstract.sec.spec}), which demonstrates formally that RGA provides SEC.

\section{Two other CRDTs: Counter and Set}
\label{sect.simple.crdts}

To demonstrate that our proof framework provides reusable components that significantly simplify SEC proofs for new algorithms, we show proofs for two other well-known operation-based CRDTs: the Observed-Remove Set (ORSet) and the Increment-Decrement Counter as described by \citet{Shapiro:2011wy}.
These proofs build upon the abstract convergence theorem and the network model of Sections~\ref{sect.abstract.convergence} and~\ref{sect.network}, and reuse some of the proof techniques developed in the formalisation of RGA in Section~\ref{sect.rga}.

As these proofs leverage the framework's machinery and proof techniques, we were able to develop them very quickly: the counter was proved correct in a matter of minutes, and the specification and correctness proof of the ORSet was done in about four hours by one of the authors, an Isabelle novice who had never used any proof assistant software prior to the start of this project.
Although these anecdotes do not constitute a formal evaluation of ease of use, we take them as being an encouraging sign.

\subsection{Increment-Decrement Counter}
\label{subsect.increment-decrement.counter}

The Increment-Decrement Counter is perhaps the simplest CRDT, and a paradigmatic example of a replicated data structure with commutative operations.
As the name suggests, the data structure supports two operations: $\isa{increment}$ and $\isa{decrement}$ which respectively increment and decrement a shared integer counter:
\begin{isabelle}
\isacommand{datatype}\ operation {\isacharequal}\ Increment\ {\isacharbar}\ Decrement
\end{isabelle}
\noindent The interpretation function for these two operations is straightforward:
\begin{isabelle}
~~~~{\isachardoublequoteopen}counter{\isacharunderscore}op\ Decrement\ \=\kill
\isacommand{fun}\ counter{\isacharunderscore}op\ {\isacharcolon}{\isacharcolon}\ {\isachardoublequoteopen}operation\ {\isasymRightarrow}\ int\ {\isasymRightarrow}\ int\ option{\isachardoublequoteclose}\ \isakeyword{where}\\
~~~~{\isachardoublequoteopen}counter{\isacharunderscore}op\ Increment\>x\ {\isacharequal}\ Some\ {\isacharparenleft}x\ {\isacharplus}\ {\isadigit{1}}{\isacharparenright}{\isachardoublequoteclose}\ {\isacharbar}\\
~~~~{\isachardoublequoteopen}counter{\isacharunderscore}op\ Decrement\>x\ {\isacharequal}\ Some\ {\isacharparenleft}x\ {\isacharminus}\ {\isadigit{1}}{\isacharparenright}{\isachardoublequoteclose}
\end{isabelle}
Note that the operations do not fail on under- or overflow, as they are defined on a type of unbounded (mathematical) integers.
We could also have implemented the counter using fixed-size integers---e.g. signed 32- or 64-bit machine words---with wrap-around on overflow, which would not have impacted the proofs.
Showing commutativity of the operations is an easy exercise in applying Isabelle's proof automation:
\begin{isabelle}
\isacommand{lemma}\ {\isachardoublequoteopen}counter{\isacharunderscore}op\ x\ {\isasymrhd}\ counter{\isacharunderscore}op\ y\ {\isacharequal}\ counter{\isacharunderscore}op\ y\ {\isasymrhd}\ counter{\isacharunderscore}op\ x{\isachardoublequoteclose}
\end{isabelle}
Unlike more complex CRDTs such as RGA, the operations of the increment-decrement counter commute unconditionally.
As a result, this CRDT converges in any asynchronous broadcast network, without requiring causally ordered delivery.
For simplicity, we define $\isa{counter}$ as a simple extension of our existing $\isa{network}{\isacharunderscore}\isa{with}{\isacharunderscore}\isa{ops}$ locale.
We need only specify the interpretation function and the initial state 0:
\begin{isabelle}
\isacommand{locale}\ counter\ {\isacharequal}\ network{\isacharunderscore}with{\isacharunderscore}ops\ {\isacharunderscore}\ counter{\isacharunderscore}op\ {\isadigit{0}}
\end{isabelle}
It is then straightforward to prove that $\isa{counter}$ is a sublocale of $\isa{strong-eventual-consistency}$ (see Section~\ref{sect.abstract.sec.spec}), from which we obtain concrete convergence and progress theorems for the counter CRDT.

\subsection{Observed-Remove Set}
\label{subsect.orset}

The Observed-Remove Set (ORSet) is a well-known CRDT for implementing replicated sets, supporting two operations: \emph{adding} and \emph{removing} arbitrary elements in the set.
It has mostly been studied in its state-based formulation \cite{Bieniusa:2012wu,Bieniusa:2012gt,Brown:2014hs,Zeller:2014fl}, but here we use the operation-based formulation as described by \citet{Shapiro:2011wy}.
The name derives from the fact that the algorithm ``observes'' the state of a node when removing an element from the set, as explained below.

We start by defining the two possible operations of the datatype:
\begin{isabelle}
\isacommand{datatype}\ {\isacharparenleft}{\isacharprime}id{\isacharcomma}\ {\isacharprime}a{\isacharparenright}\ operation\ {\isacharequal}\ Add\ {\isachardoublequoteopen}{\isacharprime}id{\isachardoublequoteclose}\ {\isachardoublequoteopen}{\isacharprime}a{\isachardoublequoteclose}\ {\isacharbar}\ Rem\ {\isachardoublequoteopen}{\isacharparenleft}{\isacharprime}id\ set{\isacharparenright}{\isachardoublequoteclose}\ {\isachardoublequoteopen}{\isacharprime}a{\isachardoublequoteclose}
\end{isabelle}
\noindent Here, $\isacharprime\isa{id}$ is an abstract type of message identifiers, and the type variable $\isacharprime\isa{a}$ represents the type of values that the application wishes to add to the set.
When an element $\isa{e}$ is added to the set, the operation $\isa{Add}\ i\ e$ is tagged with a unique identifier $\isa{i}$ in order to distinguish it from other operations that may concurrently add the same element $\isa{e}$ to the set.
When an element $\isa{e}$ is removed from the set, the operation $\isa{Rem}\ is\ e$ contains a set of identifiers $\isa{is}$, identifying all of the additions of that element that causally happened-before the removal.

The state maintained at each node is a function that maps each element $\isacharprime\isa{a}$ to the set of identifiers of operations that have added that element:
\begin{isabelle}
\isacommand{type{\isacharunderscore}synonym}\ {\isacharparenleft}{\isacharprime}id{\isacharcomma}\ {\isacharprime}a{\isacharparenright}\ state\ {\isacharequal}\ {\isachardoublequoteopen}{\isacharprime}a\ {\isasymRightarrow}\ {\isacharprime}id\ set{\isachardoublequoteclose}
\end{isabelle}
We consider an element $\isacharprime\isa{a}$ to be a member of the ORSet if the set of addition identifiers is non-empty.
The initial state of a node---the empty ORSet---is then simply $\isasymlambda\isa{x}\isachardot\ \isacharbraceleft\isacharbraceright$, i.e. the function that maps every possible element $\isacharprime\isa{a}$ to the empty set of identifiers $\isacharbraceleft\isacharbraceright$.

When interpreting an $\isa{Add}$ operation, we must add the identifier of that operation to the node state.
When interpreting a $\isa{Rem}$ operation, we must update the node state to remove all causally prior $\isa{Add}$ identifiers.
If there are no concurrent additions of the same element, this has the effect of making the set of identifiers for that element empty, and thus considering the element as no longer being in the set.
We express this as follows:
\begin{isabelle}
~~~~~~~~let\ \=before\ \={\isacharequal}\ case\ oper\ of\ \=Rem\ \=is\ \=e\kill
\isacommand{definition}\ op{\isacharunderscore}elem\ {\isacharcolon}{\isacharcolon}\ {\isachardoublequoteopen}{\isacharparenleft}{\isacharprime}id{\isacharcomma}\ {\isacharprime}a{\isacharparenright}\ operation\ {\isasymRightarrow}\ {\isacharprime}a{\isachardoublequoteclose}\ \isakeyword{where}\\
~~~~{\isachardoublequoteopen}op{\isacharunderscore}elem\ oper\ {\isasymequiv}\ case\ oper\ of\ Add\ i\ e\ {\isasymRightarrow}\ e\ {\isacharbar}\ Rem\ is\ e\ {\isasymRightarrow}\ e{\isachardoublequoteclose}\\[4pt]
\isacommand{definition}\ interpret{\isacharunderscore}op\ {\isacharcolon}{\isacharcolon}\ {\isachardoublequoteopen}{\isacharparenleft}{\isacharprime}id{\isacharcomma}\ {\isacharprime}a{\isacharparenright}\ operation\ {\isasymRightarrow}\ {\isacharparenleft}{\isacharprime}id{\isacharcomma}\ {\isacharprime}a{\isacharparenright}\ state\ {\isasymRightarrow}\ {\isacharparenleft}{\isacharprime}id{\isacharcomma}\ {\isacharprime}a{\isacharparenright}\ state\ option{\isachardoublequoteclose}\ \isakeyword{where}\\
~~~~{\isachardoublequoteopen}interpret{\isacharunderscore}op\ oper\ state\ {\isasymequiv}\\
~~~~~~~~let\>before\>{\isacharequal}\ state\ {\isacharparenleft}op{\isacharunderscore}elem\ oper{\isacharparenright}{\isacharsemicolon}\\
\>after\>{\isacharequal}\ case\ oper\ of \>Add\>i\>e\ {\isasymRightarrow}\ before\ {\isasymunion}\ {\isacharbraceleft}i{\isacharbraceright}\ {\isacharbar}\\
\>\>\>Rem\>is\>e\ {\isasymRightarrow}\ before\ {\isacharminus}\ is\\
~~~~~~~~in \>Some\ {\isacharparenleft}state\ {\isacharparenleft}{\isacharparenleft}op{\isacharunderscore}elem\ oper{\isacharparenright}\ {\isacharcolon}{\isacharequal}\ after{\isacharparenright}{\isacharparenright}{\isachardoublequoteclose}
\end{isabelle}
Here, $\isa{state}{\isacharparenleft}{\isacharparenleft}op{\isacharunderscore}elem\ oper{\isacharparenright}\ {\isacharcolon}{\isacharequal}\ \isa{after}{\isacharparenright}$ is Isabelle's syntax for pointwise function update.
A remove operation effectively undoes the prior additions of that element of the set, while leaving any concurrent or later additions of the same element unaffected.
When an element $e$ is concurrently added and removed, the identifier of the addition operation will not be in the identifier set of the removal operation.
As a result, the final state after interpreting these two operations will contain the element $e$.

As the last part of specifying ORSet, we must require that $\isa{Add}$ and $\isa{Rem}$ use identifiers correctly.
We require the identifier of $\isa{Add}$ operations to be globally unique, which we can express by making it equal to the unique ID of the message containing the operation (Section~\ref{sect.network.broadcast}).
A $\isa{Rem}$ operation must contain the set of addition identifiers in the node state at the moment when the $\isa{Rem}$ operation was issued.
We express these constraints using the following $\isa{valid-behaviours}$ predicate:
\begin{isabelle}
~~~~~~~~case\ msg\ of\ \={\isacharparenleft}i{\isacharcomma}\ Rem\ \=is\ \=e\kill
\isacommand{definition}\ valid{\isacharunderscore}behaviours\ {\isacharcolon}{\isacharcolon}\ {\isachardoublequoteopen}{\isacharparenleft}{\isacharprime}id{\isacharcomma}\ {\isacharprime}a{\isacharparenright}\ state\ {\isasymRightarrow}\ {\isacharprime}id\ {\isasymtimes}\ {\isacharparenleft}{\isacharprime}id{\isacharcomma}\ {\isacharprime}a{\isacharparenright}\ operation\ {\isasymRightarrow}\ bool{\isachardoublequoteclose}\ \isakeyword{where}\\
~~~~{\isachardoublequoteopen}valid{\isacharunderscore}behaviours\ state\ msg\ {\isasymequiv}\\
~~~~~~~~case\ msg\ of \>{\isacharparenleft}i{\isacharcomma}\ Add\>j\>e{\isacharparenright}\ {\isasymRightarrow}\ i\ {\isacharequal}\ j\ {\isacharbar}\\
\>{\isacharparenleft}i{\isacharcomma}\ Rem\>is\>e{\isacharparenright}\ {\isasymRightarrow}\ is\ {\isacharequal}\ state\ e{\isachardoublequoteclose}
\end{isabelle}
To prove that ORSet satisfies the specification of strong eventual consistency, we follow the same pattern as before.
We first define a locale $\isa{orset}$ that extends $\isa{network-with-constrained-ops}$:
\begin{isabelle}
\isacommand{locale}\ orset\ {\isacharequal}\ network{\isacharunderscore}with{\isacharunderscore}constrained{\isacharunderscore}ops\ {\isacharunderscore}\ interpret{\isacharunderscore}op\ {\isachardoublequoteopen}{\isacharparenleft}{\isasymlambda}x{\isachardot}\ {\isacharbraceleft}{\isacharbraceright}{\isacharparenright}{\isachardoublequoteclose}\ valid{\isacharunderscore}behaviours
\end{isabelle}

\noindent 
Recall the requirements of the $\isa{strong-eventual-consistency}$ specification (Section~\ref{sect.abstract.sec.spec}).
Firstly, we must show that $\isa{apply-operations}$ never fails, which is easy in this case, since the interpretation function never returns $\isa{None}$:
\begin{isabelle}
\isacommand{theorem}\ apply{\isacharunderscore}operations{\isacharunderscore}never{\isacharunderscore}fails{\isacharcolon}\\
~~~~\isakeyword{shows}\ {\isachardoublequoteopen}hb.apply{\isacharunderscore}operations\ {\isacharparenleft}node{\isacharunderscore}deliver{\isacharunderscore}messages\ {\isacharparenleft}history\ i{\isacharparenright}{\isacharparenright}\ {\isasymnoteq}\ None{\isachardoublequoteclose}
\end{isabelle}
\noindent Secondly, we must show that concurrent operations commute.
Isabelle's proof automation can easily verify that two addition operations commute unconditionally, as do two removal operations:
\begin{isabelle}
\isacommand{lemma}\ add{\isacharunderscore}add{\isacharunderscore}commute{\isacharcolon}\\
~~~~\isakeyword{shows}\ {\isachardoublequoteopen}{\isasymlangle}Add\ i{\isadigit{1}}\ e{\isadigit{1}}{\isasymrangle}\ {\isasymrhd}\ {\isasymlangle}Add\ i{\isadigit{2}}\ e{\isadigit{2}}{\isasymrangle}\ {\isacharequal}\ {\isasymlangle}Add\ i{\isadigit{2}}\ e{\isadigit{2}}{\isasymrangle}\ {\isasymrhd}\ {\isasymlangle}Add\ i{\isadigit{1}}\ e{\isadigit{1}}{\isasymrangle}{\isachardoublequoteclose}\\[4pt]
\isacommand{lemma}\ rem{\isacharunderscore}rem{\isacharunderscore}commute{\isacharcolon}\\
~~~~\isakeyword{shows}\ {\isachardoublequoteopen}{\isasymlangle}Rem\ i{\isadigit{1}}\ e{\isadigit{1}}{\isasymrangle}\ {\isasymrhd}\ {\isasymlangle}Rem\ i{\isadigit{2}}\ e{\isadigit{2}}{\isasymrangle}\ {\isacharequal}\ {\isasymlangle}Rem\ i{\isadigit{2}}\ e{\isadigit{2}}{\isasymrangle}\ {\isasymrhd}\ {\isasymlangle}Rem\ i{\isadigit{1}}\ e{\isadigit{1}}{\isasymrangle}{\isachardoublequoteclose}
\end{isabelle}
\noindent However, add and remove operations commute only if the identifier of the addition is not one of the identifiers affected by the removal:
\begin{isabelle}
~~~~\isakeyword{assumes}\ \=\kill
\isacommand{lemma}\ add{\isacharunderscore}rem{\isacharunderscore}commute{\isacharcolon}\\
~~~~\isakeyword{assumes}\>{\isachardoublequoteopen}i\ {\isasymnotin}\ is{\isachardoublequoteclose}\\
~~~~\isakeyword{shows}\>{\isachardoublequoteopen}{\isasymlangle}Add\ i\ e{\isadigit{1}}{\isasymrangle}\ {\isasymrhd}\ {\isasymlangle}Rem\ is\ e{\isadigit{2}}{\isasymrangle}\ {\isacharequal}\ {\isasymlangle}Rem\ is\ e{\isadigit{2}}{\isasymrangle}\ {\isasymrhd}\ {\isasymlangle}Add\ i\ e{\isadigit{1}}{\isasymrangle}{\isachardoublequoteclose}
\end{isabelle}

Proving that the assumption $\isa{i} \mathbin{\isasymnotin} \isa{is}$ holds for all concurrent $\isa{Add}$ and $\isa{Rem}$ operations is a bit more laborious.
We define \isa{added-ids} to be the identifiers of all $\isa{Add}$ operations in a list of delivery events, even if those elements are subsequently removed.
Then we prove that the set of identifiers in the node state is a subset of \isa{added-ids} (since $\isa{Add}$ operations only ever add identifiers to the node state, and $\isa{Rem}$ operations only ever remove identifiers):
\begin{isabelle}
~~~~\isakeyword{assumes}\ \=\kill
\isacommand{lemma}\ apply{\isacharunderscore}operations{\isacharunderscore}added{\isacharunderscore}ids{\isacharcolon}\\
~~~~\isakeyword{assumes}\>{\isachardoublequoteopen}{\isasymexists}suf{\isachardot}\ pre\ {\isacharat}\ suf\ {\isacharequal}\ history\ i{\isachardoublequoteclose}\\
~~~~~~~~\isakeyword{and}\>{\isachardoublequoteopen}apply{\isacharunderscore}operations\ pre\ {\isacharequal}\ Some\ state{\isachardoublequoteclose}\\
~~~~\isakeyword{shows}\>{\isachardoublequoteopen}state\ e\ {\isasymsubseteq}\ set\ {\isacharparenleft}added{\isacharunderscore}ids\ pre\ e{\isacharparenright}{\isachardoublequoteclose}
\end{isabelle}
\noindent From this lemma, we deduce that when an $\isa{Add}$ and a $\isa{Rem}$ operation are concurrent, the identifier of the $\isa{Add}$ cannot be in the set of identifiers removed by $\isa{Rem}$:
\begin{isabelle}
~~~~\isakeyword{assumes}\ \={\isachardoublequoteopen}Rem\ \=is\ \=e{\isadigit{2}}\ \={\isasymin}\kill
\isacommand{lemma}\ concurrent{\isacharunderscore}add{\isacharunderscore}remove{\isacharunderscore}independent{\isacharcolon}\\
~~~~\isakeyword{assumes}\>{\isachardoublequoteopen}{\isacharparenleft}Add\ i\ e{\isadigit{1}}{\isacharparenright}\ $\|$ {\isacharparenleft}Rem\ is\ e{\isadigit{2}}{\isacharparenright}{\isachardoublequoteclose}\ \\
~~~~~~~~\isakeyword{and}\>{\isachardoublequoteopen}Add\>i\>e{\isadigit{1}}\>{\isasymin}\ set\ {\isacharparenleft}node{\isacharunderscore}deliver{\isacharunderscore}messages\ {\isacharparenleft}history\ j{\isacharparenright}{\isacharparenright}{\isachardoublequoteclose}\\
~~~~~~~~\isakeyword{and}\>{\isachardoublequoteopen}Rem\>is\>e{\isadigit{2}}\>{\isasymin}\ set\ {\isacharparenleft}node{\isacharunderscore}deliver{\isacharunderscore}messages\ {\isacharparenleft}history\ j{\isacharparenright}{\isacharparenright}{\isachardoublequoteclose}\\
~~~~\isakeyword{shows}\>{\isachardoublequoteopen}i\ {\isasymnotin}\ is{\isachardoublequoteclose}
\end{isabelle}
\noindent Now that we have proved that the assumption of $\isa{add-rem-commute}$ holds for all concurrent operations, we can deduce that all concurrent operations commute:
\begin{isabelle}
\isacommand{theorem}\ concurrent{\isacharunderscore}operations{\isacharunderscore}commute{\isacharcolon}\\
~~~~\isakeyword{shows}\ \ {\isachardoublequoteopen}hb{\isachardot}concurrent{\isacharunderscore}ops{\isacharunderscore}commute\ {\isacharparenleft}node{\isacharunderscore}deliver{\isacharunderscore}messages\ {\isacharparenleft}history\ i{\isacharparenright}{\isacharparenright}{\isachardoublequoteclose}
\end{isabelle}

Having proved $\isa{apply-operations-never-fails}$ and $\isa{concurrent-operations-commute}$, we can now immediately prove that $\isa{orset}$ is a sublocale of $\isa{strong-eventual-consistency}$, using the familiar proof pattern from the other CRDTs.
This proof produces concrete convergence and progress theorems for the ORSet.

\section{Related work}\label{sect.relatedwork}

In a system where different nodes may concurrently perform updates without coordinating with each other, strong eventual consistency requires a conflict resolution algorithm to reconcile concurrent updates. 
In some cases, a trivial algorithm is used, for example:
\begin{description}
\item[User-defined conflict resolution:] Some systems store all conflicting versions of the data,
and either leave it for manual resolution by a user, or invoke a user-defined merge function.
However, manual resolution is an unacceptable burden for the user in many applications, and defining
merge functions in application code is error-prone; for example, \citet{DeCandia:2007ui} describe a
shopping cart anomaly at Amazon that arose due to poor conflict resolution.

\item[Last write wins (LWW):] Each version of the data structure is assigned a unique timestamp.
When there is a conflict, the system picks the version with the highest timestamp and discards other
versions. Apache Cassandra takes this approach, for example \cite{KingsburyCassandra}.
Although LWW achieves convergence, it does so at the cost of losing user input, which is often unacceptable.
\end{description}
However, there are also algorithms that achieve convergence automatically, without discarding updates.
In Section~\ref{sect.related.ot.crdts} we summarise two main lines of work, CRDTs and OT, which have the same fundamental goal of conflict resolution and convergence, but which take different approaches towards achieving it.
In Section~\ref{sect.related.verification} we discuss existing work on formal verification of those algorithms.

\subsection{Operational Transformation (OT) and Conflict-free Replicated Data Types (CRDTs)}\label{sect.related.ot.crdts}

Algorithms for achieving strong eventual consistency have been studied extensively in the context of collaborative editing and groupware.
The \emph{operational transformation} (OT) approach was developed to allow several users to concurrently modify a document, applying edits immediately to their local copy, propagating them asynchronously to other users, and automatically resolving any conflicts such that all nodes converge towards the same state.

OT algorithms for text documents include dOPT \cite{Ellis:1989ue}, Jupiter \cite{Nichols:1995fd}, adOPTed \cite{Ressel:1996wx}, GOT \cite{Sun:1998un}, GOTO \cite{Sun:1998vf}, SOCT2 \cite{Suleiman:1997gl,Suleiman:1998eu}, SOCT3/4 \cite{Vidot:2000ch}, IMOR \cite{Imine:2003ks}, SDT \cite{Li:2004er,Li:2008hw}, and TTF \cite{Oster:2006tr}.
The approach has also been generalised to other data structures such as XML trees \cite{Ignat:2003jy,Davis:2002iv,Jungnickel:2015ua} and vector graphics documents \cite{Sun:2002jb}.

Many OT algorithms assume that operations are sequenced through a central server and delivered to all clients in the same order.
This design was originally pioneered by the Jupiter system \cite{Nichols:1995fd} and is now used by all widely-deployed OT-based collaboration systems, including Google Docs \cite{DayRichter:2010tt}, Microsoft Word Online, Etherpad \cite{Etherpad:2011um}, Google Wave/Apache Wave \cite{Wang:2015vo}, and Novell Vibe \cite{Spiewak:2010vw}.

OT algorithms track the version of the document in which each operation applies, and if an operation needs to be applied to a later document version (because another, concurrent operation has already been applied), the operation must be transformed.
\citet{Ressel:1996wx} introduced two properties that the OT transformation function must satisfy, which are known as $\mathit{TP}_1$ and $\mathit{TP}_2$.

Given two concurrent operations $x$ and $y$ that modify the same initial state, $\mathit{TP}_1$ requires that $y$ can be transformed into an operation $y'$ that performs an equivalent modification on a state where $x$ has already been applied, and vice versa, such that $x \circ y' = y \circ x'$.
Systems that sequence operations through a central server need only satisfy $\mathit{TP}_1$ because each client only needs to reorder its operations with respect to the server's operation sequence.

However, as discussed in the introduction, we are interested in replication algorithms for decentralised systems without any central server.
If there are three concurrent operations $x$, $y$, and $z$ that modify the same initial state, and those operations can be applied in any order, $\mathit{TP}_1$ does not suffice, and the $\mathit{TP}_2$ property must also be satisfied.
$\mathit{TP}_2$ requires that if transformations of $x$ and $y$ are applied in either order, the same transformation of $z$ can be applied to the result: $x \circ y' \circ z' = y \circ x' \circ z'$.
Since transformed operations may be different from original operations, this property demands much more than just commutativity, making it difficult to implement correctly.
We show in Section~\ref{sect.related.verification} how almost all OT algorithms have failed to satisfy $\mathit{TP}_2$.

Instead, \emph{conflict-free replicated data types} (CRDTs) have been developed to achieve SEC in decentralised systems.
As we have noted, CRDTs make operations commutative by design by attaching additional metadata to the data structure.
To propagate changes between nodes, a CRDT either captures every update as an operation and broadcasts it to other nodes (an \emph{operation-based} CRDT), or periodically broadcasts its entire node state (a \emph{state-based} CRDT).
Operation-based CRDTs require operations to commute; state-based CRDTs require a merge function over a join-semilattice, allowing two states to be combined such that the result reflects changes made in both nodes \cite{Shapiro:2011wy,Shapiro:2011un}.
State-based CRDTs have been deployed commercially in the Riak database \cite{Brown:2014hs}, but in this work we focus on operation-based algorithms, because all known CRDTs for text editing and ordered lists are operation-based.

As with OT, several CRDTs for text documents have been developed, including RGA \cite{Roh:2011dw}, Treedoc \cite{Preguica:2009fz}, WOOT \cite{Oster:2006wj}, Logoot \cite{Weiss:2010hx}, and LSEQ \cite{Nedelec:2013ky,Nedelec:2016eo}.
Other datatypes include registers and counters \cite{Shapiro:2011wy,Shapiro:2011un}, maps \cite{Baquero:2016iv}, sets \cite{Bieniusa:2012wu,Bieniusa:2012gt}, XML \cite{Martin:2010ih}, and JSON trees \cite{Kleppmann:2016ve}.
Cloud types \cite{Burckhardt:2012jy} have similarities to CRDTs, using a relational data model.

\subsection{Formal verification}\label{sect.related.verification}

The history of algorithms for achieving convergence in a distributed setting has been fraught with difficulty.
Informal reasoning has repeatedly produced approaches that fail to converge in certain scenarios, and even several formal ``proofs'' later turned out to be false, as explained below.
For OT, as described in Section~\ref{sect.related.ot.crdts}, convergence in this setting requires satisfying the $\mathit{TP}_1$ and $\mathit{TP}_2$ properties.
While $\mathit{TP}_1$ has proved to be readily achievable in practice, and all the aforementioned widely-deployed OT systems rely on it, the $\mathit{TP}_2$ property has been a significant source of problems.

The original peer-reviewed publications of dOPT, adOPTed, IMOR, SOCT2, and SDT all claimed that their transformation functions satisfied $\mathit{TP}_2$, but those claims were subsequently shown to be false by giving counter-examples \cite{Imine:2003ks,Imine:2006kn,Oster:2005vi}.
In the case of dOPT and adOPTed, the $\mathit{TP}_2$ claim had originally been asserted without proof.
In the case of SOCT2 and SDT, there were hand-written ``proofs'' that later turned out to be incorrect.
For IMOR and SOCT2, there had even been machine-checked ``proofs'' \cite{Imine:2003ks}, but \citet{Oster:2005vi} showed that they were also invalid because they had made incorrect assumptions.

\citet{Randolph:2015gj} have even shown that in the classic formulation of OT it is impossible to
achieve $\mathit{TP}_2$. To our knowledge, TTF is at present the only $\mathit{TP}_2$-claiming OT
algorithm for which no counter-example is known, and it circumvents the impossibility result of
\citet{Randolph:2015gj} by using a different formulation of the transformation
\cite{Oster:2006tr,Levien:2016wz}.

Formal proofs of the $\mathit{TP}_1$ property have been more successful: \citet{Sinchuk:2016cf} use Coq to verify that their algorithm satisfies $\mathit{TP}_1$, and \citet{Jungnickel:2015ua} use Isabelle/HOL for the same purpose.
For CRDTs, the only machine-checked verification of which we are aware is an Isabelle formalisation of state-based sets, registers, and counters by \citet{Zeller:2014fl}; this work does not consider any list datatypes or any operation-based CRDTs.

The convergence of the RGA CRDT for ordered lists, which we study in this paper, has previously been
demonstrated in handwritten proofs \cite{Attiya:2016kh,Kleppmann:2016ve,Roh:2009ws}. Although we
have no reason to doubt the correctness of those proofs, the historic experience with
$\mathit{TP}_2$ makes us wary of claims whose assumptions and reasoning process have not been
checked rigorously. Other authors have also pointed out that handwritten proofs are laborious and
difficult to check by hand \cite{Li:2008hw,Li:2005jq}.

To our knowledge, our work is the first mechanised proof of operation-based CRDTs in general, and of
any ordered list CRDT in particular. As \citet{Oster:2005vi} have demonstrated, machine-checked
proofs are not immune to errors that are due to false assumptions. To avoid this trap, we prove not
only the commutativity of operations (which is subject to certain assumptions), but also that those
assumptions are guaranteed to hold in all behaviours of our network model. The network model in turn
is specified by a small set of axioms that are not specific to any particular CRDT, and whose
correctness can be robustly defended (see Section~\ref{sect.network}).

\citet{Burckhardt:2014ft} present a framework for specifying and reasoning about replicated datatypes, but do not support mechanised proofs at present, and use different techniques to those described in this paper.

More generally, applying verification techniques to distributed systems is an active area of research.
Interactive theorem provers~\cite{DBLP:conf/pldi/WilcoxWPTWEA15,DBLP:journals/afp/DebratM12,DBLP:conf/sss/Charron-BostDM11}, model checkers~\cite{DBLP:conf/asm/AzmyMW16,DBLP:journals/entcs/JohnsonLLV04}, and formal specification tools~\cite{DBLP:journals/ijaacs/TounsiMM16,DBLP:conf/asm/AndriamiarinaMS14,DBLP:conf/wetice/TounsiMM13} have all been adopted for the verification and specification of distributed systems, algorithms, and protocols.
Interestingly, recent empirical work~\cite{DBLP:conf/eurosys/FonsecaZWK17} has found that several verified distributed systems contain critical bugs that can cause runtime crashes or the return of incorrect results to clients---violating the supposed guarantees offered by their correctness theorems.
A common cause of these bugs is a mismatch between the assumptions made when verifying the system and the guarantees offered by the underlying network, libraries, and operating system infrastructure upon which they are built.
We see this as compelling evidence that verifying distributed systems starting from a model of the network and building up, as we do in this work, is a robust approach to distributed systems verification.

\section{Discussion}
\label{sect.discussion}

The convergence proofs for all of our CRDT implementations follow the same structure.
First we define the type of local state at each node, and the types of operations that may be invoked to modify the state.
When one node invokes an operation, it is broadcast to other nodes using our network model, implemented as a specialisation of the $\isa{network}{\isacharunderscore}\isa{with}{\isacharunderscore}(\isa{constrained}{\isacharunderscore})\isa{ops}$ locale.
An interpretation function is called whenever a message containing an operation is delivered to a node, and it transforms that node's local state to incorporate the operation.
To demonstrate convergence, we must show that all operations commute with themselves and with each other, subject to certain assumptions.
Next, we must prove that those assumptions are always satisfied by any concurrent operations in the network.
Finally, a CRDT must demonstrate that applying an operation never fails, provided that the operation was constructed according to the definition of the algorithm.

When these proof obligations have been met, we are able to conclude that the algorithm satisfies our abstract specification $\isa{strong-eventual-consistency}$, from which we obtain convergence and progress theorems for the replicated datatype.
The abstract specification is independent of any particular network model or replication algorithm, and we assert that it constitutes a general but precise definition of strong eventual consistency.
As this recurring pattern demonstrates, we have not only isolated reusable lemmas and models of networks, but also a proof strategy that algorithm designers can use to obtain a convergence theorem for their operation-based CRDT.

Over half of our development---the network model, convergence proof, and lemmas---is independent of any particular CRDT and is reusable in future proofs.
In particular, we use: around 620 lines for our network model, around 380 lines for the abstract convergence proof, 775 lines for the RGA proof, around 270 lines for the ORSet proof, and around 55 lines for the Counter proof.
Additional shared code consists of around 170 lines of source.
Definitions and proofs of correctness for our three CRDT implementations are pleasingly short: all three are shown to be convergent in fewer than 800 lines of source, using the proof strategy described above. 

Lastly, all three of our CRDT implementations are ``executable'' in the sense that we can use Isabelle's code generation mechanism to obtain OCaml (or Scala, SML, and Haskell) implementations from our definitions~\cite{DBLP:conf/flops/HaftmannN10}.
We have run an extraction of one of our CRDTs---the counter---on a simple network of $n$ nodes, communicating over TCP links between machines.
The purpose of this extraction is to demonstrate that we have not used any uncomputable functions in our Isabelle definitions.
We leave a detailed empirical evaluation of the algorithms, such as tests of their performance and fault tolerance, for future work.

\section{Conclusion}
\label{sect.conclusion}

In this paper we adopted a ``foundational'' approach to proving the correctness of a class of SEC algorithms: Conflict-free Replicated Datatypes (CRDTs).
In our work, we made no axiomatic assumptions related to any individual algorithm; instead, we constructed a formal, realistic model of a computer network that may delay, drop, or reorder messages sent between computers---a model well known within the distributed systems community, with defensible axioms.
In addition, we isolated a formal specification of SEC, and showed that any algorithm that meets the preconditions of this specification must converge.
Our network model, the SEC specification, and a library of lemmas form a framework with which one can prove convergence for concrete CRDT implementations.

As a case study in applying our framework, we formalised three operation-based CRDTs---the Replicated Growable Array, the Observed-Remove Set, and an increment-decrement Counter.
For each algorithm we proved that its assumptions are satisfied in all possible network behaviours, and that it satisfies the preconditions of our abstract SEC specification, obtaining a guarantee that each of our three CRDT implementations converge.
Moreover, these convergence theorems were obtained using only a thin layer of CRDT-specific code, using a fixed pattern of proof in all three cases.
Since informal reasoning about convergent replication algorithms has been shown to difficult and error-prone, as exemplified by various failed proofs (Section~\ref{sect.related.verification}) and the ``a bit of a mystery'' RGA \cite{Attiya:2016kh}, we believe that this formal verification is particularly important.

Our work has been motivated by the desire to support a wider range of strong eventual consistency algorithms in distributed systems.
Operational Transformation algorithms based on the $\mathit{TP}_1$ property alone are widely used in systems such as Google Docs (see Section~\ref{sect.related.verification}), but these algorithms require clients to communicate all changes via a single central server.
On the other hand, state-based CRDTs are used in systems such as Riak.
Both of these approaches are limiting.
In the case of Operational Transformation, the requirement for a central server increases the risk of faults, and prevents direct collaboration between devices via local wireless network connections.
In the case of state-based CRDTs, efficient algorithms to support complex data structures such as ordered lists are yet to be found.
Consequently, our approach provides the groundwork for a new generation of applications that use truly distributed strong eventual consistency algorithms, including robust, collaborative applications working on complex data structures in a peer-to-peer setting, something which is likely to become increasingly important in a world where mobile devices are becoming increasingly prevalent.

Although we have focussed on operation-based algorithms in this work, we speculate that our framework is also amenable to formalising Operational Transformation algorithms and state-based CRDTs.
We also speculate that our framework could be used to demonstrate the equivalence of classes of strong eventual consistency algorithms.
We leave this to future work.

\begin{acks}
    The authors wish to acknowledge the support of the~\grantsponsor{GS501100000266}{EPSRC}{http://dx.doi.org/10.13039/501100000266} ``REMS: Rigorous Engineering for Mainstream Systems'' programme grant (\grantnum{GS501100000266}{EP/K008528}),
    the~\grantsponsor{GS501100000266}{EPSRC}{http://dx.doi.org/10.13039/501100000266} ``Interdisciplinary Centre for Finding, Understanding and Countering Crime in the Cloud'' grant (\grantnum{GS501100000266}{EP/M020320}),
    and \grantsponsor{GS100000003}{The Boeing Company}{http://dx.doi.org/10.13039/100000003}.
    We thank Peter Sewell, Alan Mycroft, and the anonymous reviewers for their helpful feedback on this paper.
\end{acks}


\end{document}